\documentclass[graybox]{svmult}

\usepackage{type1cm}        % activate if the above 3 fonts are
\usepackage{makeidx}         % allows index generation
\usepackage{graphicx}        % standard LaTeX graphics tool
\usepackage{multicol}        % used for the two-column index
\usepackage[bottom]{footmisc}% places footnotes at page bottom

\usepackage{newtxtext}       % 
\usepackage{newtxmath}       % selects Times Roman as basic font
\UseRawInputEncoding

\usepackage{physics}
\usepackage[amssymb,Gray]{SIunits}
\usepackage{tikz}
\usepackage[colorlinks]{hyperref}

\makeindex             % used for the subject index
\newcommand{\bleq}{\ensuremath{\mathrel{\phantom{=}}}}
\newcommand{\nnl}{\nonumber\\}
\newcommand{\up}{\uparrow}
\newcommand{\dn}{\downarrow}
\newcommand{\gspac}[1][]{{^{(3)}\!g}_{#1}} % spatial metric with optional subscript, and ...
\newcommand{\gspacInv}[1][]{{^{(3)}\!g^{-1}_{#1}}} % ... the same for its inverse
\newcommand{\supp}{\mathrm{supp}} % support
\newcommand{\schr}{Schr{\"o}\-din\-ger}
\newcommand{\QFT}{quantum field theory}
\begin{document}

\title*{Coupling Quantum Matter and Gravity}
\author{Domenico Giulini, Andr\'e Gro\ss ardt and Philip K.\ Schwartz}
\institute{Domenico Giulini \at Institute for Theoretical Physics, University of Hannover, Appelstra\ss e 2, 30167 Hannover, Germany, and Center of Applied Space Technology and Microgravity, University of Bremen, Am Fallturm 1, 28359 Bremen, Germany, \email{giulini@itp.uni-hannover.de}
\and Andr\'e Gro\ss ardt \at Institute for Theoretical Physics, Friedrich Schiller University Jena, Fr\"obelstieg 1, 07743 Jena, Germany, \email{andre.grossardt@uni-jena.de}
\and Philip K.\ Schwartz \at Institute for Theoretical Physics, University of Hannover, Appelstra\ss e 2, 30167 Hannover, Germany, \email{philip.schwartz@itp.uni-hannover.de}}
\maketitle

\vspace{-8\baselineskip}

\abstract{In this contribution we deal with 
several issues one encounters when trying to 
couple quantum matter to classical gravitational
fields. We start with a general background discussion 
and then move on to two more technical sections. 
In the first technical part we consider the 
question how the Hamiltonian of a composite 
two-particle system in an external gravitational 
field can be computed in a systematic post-Newtonian 
setting without backreaction. This enables us to 
reliably estimate the consistency and completeness of less systematic and more intuitive approaches that attempt 
to solve this problem by adding `relativistic effects' 
by hand. In the second technical part we consider the 
question of how quantum matter may act as source for 
classical gravitational fields via the semiclassical 
Einstein equations. Statements to the effect that 
this approach is fundamentally inconsistent are 
critically reviewed.}

\section{Introduction and preliminary discussion}
\label{sec:intro}
The central concern of this contribution is the relation 
between gravity---as described by the classical (i.e.\ 
not quantised) theory of General Relativity (henceforth 
abbreviated by GR)---and the theory of all other 
`interactions', which are  described by relativistic 
quantum (field) theories (henceforth abbreviated by 
RQFT).\footnote{Note that we distinguish the general 
notion of \QFT\ from the specific form it takes in presence 
of Poincar\'e invariance, in which case we write
RQFT.}
To this end, we will address several specific 
technical as well as conceptual issues which we consider
important. Some of these issues can be resolved
(we believe), whereas others may perhaps require 
rethinking before any resolution can be proposed. 

In this first section we wish to share and discuss 
a few thoughts concerning the relation 
of `gravity' on one side, and `matter' on the 
other. This preliminary discussion is not 
only meant to set the stage for the later, more 
technical parts of this contribution, but also 
tries to convey a sense of appreciation for 
the distinguishing features of the `gravitational 
interaction'. 

In section~\ref{sec:qm-class-backgrounds} we first 
show how to systematically incorporate in form of a 
post-Newtonian expansion the interaction of an 
electromagnetically bound model-atom (consisting 
of two charged point particles) with an external 
gravitational field. A second part of that section 
revisits in some detail the question of allegedly 
`anomalous' couplings of internal energies 
to the centre-of-mass motion of composite systems. 

Section~\ref{sec:qm-backreaction} deals with the 
question of backreaction, i.e.\ how quantum matter
might source a classical gravitational field. Much debated issues concerning consistency and causality are addressed
in separate subsections, as well as alternative schemes
to that of semiclassical gravity, like collapse models.

\subsection{Why care?}
\label{subsec:whyCare}
We recall Einstein's equations\footnote{%
Our conventions are the usual ones:
$R_{\mu\nu}:=R^\lambda_{\phantom{\lambda}\mu\lambda\nu}$
is the Ricci tensor if $R^\lambda_{\phantom{\lambda}\mu\sigma\nu}$
denotes the Riemann tensor, $R:=g^{\mu\nu}R_{\mu\nu}$
is the Ricci scalar, $T_{\mu\nu}$ denote the covariant
components of the energy-momentum tensor with 
$T_{00}:=T(e_0,e_0)$ the energy-density for the 
observer characterised by the unit timelike 
vector $e_0$. $G$ is Newton's gravitational constant 
and $c$ the speed of light in vacuum. Our signature
convention is `mostly plus' $(-,+,+,+)$.}
\begin{equation}
\label{eqn:einstein-eq}
        R_{\mu\nu} - \frac{1}{2} R \, g_{\mu\nu} = \frac{8 \pi G}{c^4} T_{\mu\nu} \,,
    \end{equation}
which relate the spacetime metric $g$  to the 
matter content,
though the matter does not determine the metric. 
More precisely, the ten components $T_{\mu\nu}$ 
comprising the matter's energy and momentum 
densities and flux-densities,  
determine the 10 Ricci components out of the 
20 Riemann curvature components. The metric carries 
its own degrees of freedom, over and above those 
of the matter, which are capable to transport 
physical quantities like energy and momentum 
from one material system to another through 
spacetime regions which are entirely devoid of 
any matter. This happens, for example, if a distant 
binary star-system emits gravitational waves to 
a detector on Earth. According to GR, the mere existence of spacetime is 
logically independent of the existence of matter. 
Hence we arrive at the following dichotomy of our
fundamental laws:
\begin{itemize}
\item 
Gravity is modelled by \emph{classical} GR. It requires a spacetime consisting 
of a pair $(\mathcal{M},g)$, where $\mathcal{M}$
is a 4-dimensional differentiable manifold 
and $g$ denotes a Lorentzian metric on it. The latter 
determines a connection, the Levi--Civita connection
for $g$, and hence an inertial structure (see below). 
The coupling between spacetime and matter---sometimes
referred to as \emph{backreaction}---is then given 
by Einstein's equations~\eqref{eqn:einstein-eq}.
Classical matter is described by fields \emph{on}  
$\mathcal{M}$ (i.e.\ sections in various vector bundles 
over $\mathcal{M}$) the dynamics of which usually 
follows from a Lagrange density $\mathcal{L}$
from which we also obtain the energy-momentum tensor 
$T_{\mu\nu}$ through the functional derivative 
$\delta \mathcal{L}/\delta g^{\mu\nu}$.
\item 
Fundamentally, matter is modelled by \emph{quantum} 
field theories. In a Poincar{\'e} invariant context, 
states of quantum fields are commonly defined as 
elements of the (bosonic or fermionic) Fock 
space $\mathcal{F}_\pm(\mathcal{H})$ over a 
single-particle Hilbert space $\mathcal{H}$, usually 
(and loosely 
speaking) the space of positive-frequency solutions of 
some (classical) field equation \emph{on flat Minkowski spacetime}. 
The dynamical laws of these fields are determined by their 
interactions with each other as defined by the total Lagrange 
density $\mathcal{L}$---leading to the common picture where forces 
between fermionic particles are carried by gauge bosons.
\end{itemize}
Attempts to cure this division at the level of the 
most fundamental theories by replacing the classical spacetime 
structure and Einstein's equations by concepts more 
compatible with the Hilbert space structure of matter 
are summarised under the label of `quantum gravity' and 
have so far not led to any generally accepted scheme.  
Less pervasive `escape strategies' are those in which 
gravity stays classical. Treating gravity and matter 
with different mathematical principles need not 
necessarily result in inconsistencies. Two aspects 
of such strategies should be distinguished: 
\begin{enumerate}
\item \emph{Ignoring backreaction} means to take a fixed,
generally \emph{curved}, spacetime $(\mathcal{M},g)$ 
and consider quantum fields on it. 
At least in the regime in which the energy-momentum density of the fields is small enough compared to that of the \emph{sources} of the background gravitational field, this approach will provide a sensible approximation.
The difficult 
task remains to formulate \QFT\ without Poincar{\'e}
invariance (or any other maximal symmetry, like
de\,Sitter or anti-de\,Sitter). In spite of much 
mathematical progress over the last decades%
~\cite{waldQuantumFieldTheory1994,barQuantumFieldTheory2009,brunettiAdvancesAlgebraicQuantum2015}, 
which also shows how many of the familiar concepts 
of RQFT fail to exist in Quantum Field Theory in 
Curved Spacetime (henceforth abbreviated by QFTCS) 
due to the lack of Poincar{\'e} invariance, 
QFTCS is still far from being able to address 
and answer specific physical problems with the 
desired certainty.\footnote
{The furthest developed mathematically rigorous approach to 
QFTCS uses methods from algebraic \QFT, and has led, for 
example, to the notion of \emph{locally covariant quantum 
field theories} on curved spacetimes. Among other results, 
this allows to transfer (properly formulated versions of) 
the spin-statistics and CPT theorems from RQFT to QFTCS 
(see, in particular, chapter 4 of \cite{brunettiAdvancesAlgebraicQuantum2015}). 
As is well-known, QFTCS is also of fundamental importance 
for the physics of black holes, leading to the emission of 
Hawking radiation \cite{hawkingParticleCreation1975}, and 
is believed to play an important role in relativistic cosmology~\cite{MukhanovIntroductionQuantumEffectsGravity2007}.}

However, when considering weak gravitational fields, i.e.\ 
small deviations from Minkowskian geometry, and small 
velocities of the systems involved, we may restrict ourselves 
to fixed-particle sectors. The systems are then effectively 
described by the corresponding minimally coupled relativistic 
wave equations, and a post-Newtonian expansion of these 
equations in the parameter $c^{-1}$ yields a systematic 
post-Newtonian description of quantum systems in 
gravitational fields. In section~\ref{sec:qm-class-backgrounds}, 
we will explore this strategy in detail and discuss related issues.

\item \emph{Inclusion of backreaction} in a 
`classical-quantum scheme' must mean to consider 
the matter's energy-momentum tensor $T_{\mu\nu}$ 
on the right-hand side of the classical 
equation~\eqref{eqn:einstein-eq} as a classical 
field, like, e.g., the expectation value of the
energy-momentum tensor in some state. In that 
case, the metric will acquire a dependence on that 
state, which also re-couples to the evolution of the 
state via the matter equations (which contain 
the metric). This will effectively cause a 
non-linearity in the dynamical evolution of the 
quantum state. 

Semiclassical models of this kind are widely 
believed to be inconsistent 
and/or unphysical. However, when making 
statements along this line, one needs to be particularly 
careful, and such claims of inconsistency often turn 
out to be founded in implicit assumptions. An important aspect is the precise theoretical description of the reduction of quantum states upon measurement (or, rather, the lack thereof). In 
section~\ref{sec:qm-backreaction}, we will discuss these 
and related issues concerning the gravitational backreaction of quantum matter on a classical spacetime.
\end{enumerate}

\subsection{A note on the equivalence principle}
The equivalence principle is commonly viewed as 
the core assumption of GR. In the proper sense of the word, 
it is a \emph{heuristic} principle, characterising 
the coupling of gravity to any sort of (classical) 
matter, the dynamical laws of which have already 
been formulated in a way compatible with Special 
Relativity, i.e. in a Poincar{\'e} invariant form. 
In connection with matter described by ordinary 
Quantum Mechanics (henceforth abbreviated QM) this 
principle loses its heuristic power, one 
obvious reason being the lack of Poincar{\'e} 
invariance of the \schr\ equation. 
Speculations of whether QM violates the equivalence 
principle \emph{per se} have a long history 
and are still ongoing. In that situation we find
it useful to reflect on the meaning of this 
principle. 

The perhaps most concise way to state the implication 
of the equivalence principle, as it is realised in GR, 
is to say that the coupling to gravity is exclusively 
furnished via \emph{one and the same geometry} that 
is common to all matter components. It does 
\emph{not} say, e.g., that all `pieces of matter' 
fall alike (in a way only depending on the initial 
conditions but not on the inner composition and 
nature of the `piece'). This latter statement is 
only true for the highly idealised concept of a 
`test particle', the degree of approximate 
realisation of which strongly depends on the physical 
context. 

Indeed, a `test particle' must have an extent 
much less than the curvature radius of the ambient 
spacetime, for otherwise it cannot probe the 
\emph{local} gravitational field. On the other 
hand, its extent must be much larger than its own 
gravitational radius $GM/c^2$, for otherwise its 
own gravity dominates the ambient one and also 
its binding energy becomes of the order of its 
rest energy. Furthermore it may not have any 
appreciable spin angular momentum, higher 
mass-multipole moment (quadrupole and higher), 
charge, etc. We emphasise that such qualities 
need not disappear with vanishing size 
\cite{ohanianEP1977}. We leave it to the reader 
to come up with those physical properties that 
may still be varied within the set of `test particles'.

Hence, according to GR, the centre of mass 
of a banana will not describe the same worldline as 
that of an apple for the same initial conditions, 
due to the higher quadrupole moment of the former,
which couples to the spacetime curvature. 
Adding spin angular momentum with respect to the 
centre of mass will again influence its trajectory,
again due to curvature couplings.%
\footnote{The spin-curvature coupling is already 
present in the standard (lowest order) pole-dipole approximation of the Mathisson--Papapetrou--Dixon 
equation, the quadrupole coupling appears in the 
next order; see, e.g., the beautiful review   
\cite{dixonNewMechanicsMathisson2015}.} 
Nothing of that sort is, of course, in violation 
of the equivalence principle, since all these 
deviations can be fully accounted for by a single 
spacetime geometry. 

Likewise, in QM, the `centre' of a wave packet need 
not fall on the same trajectory as a point particle 
for equivalent initial conditions. Also, the wave 
behaviour of the former does depend on the inertial 
mass of the particle it represents (as the de~Broglie
wavelength does). Again, this does not \emph{per se} 
contradict the equivalence principle, as sometimes 
suggested following a famous 
argument by Salecker (and contradicted by Feynman) 
made during the famous 1957 Chapel Hill conference~\cite[chapter\,23]{chapelhill1957}.  In fact, 
in can be proven that \emph{any} solution to the 
\schr\ equation in a homogeneous (though arbitrarily 
time dependent) force field is obtained from a solution 
of the force-free equation by translating it along the 
integral flow of a classical solution curve and 
multiplying it with an appropriate space- and 
time-dependent phase factor~\cite{giuliniEP2012}. This 
implies that the spatial 
probability distribution falls exactly like a continuous
dust cloud of classical particles with identical 
initial velocities.
If the particles' spatial paths only depend on the ratio 
of the gravitational to the inertial mass, then so does 
the `path' of the probability distribution. Note that 
this statement is exact and holds for \emph{all} solutions 
to the \schr\ equation, not just those 
approximating classical behaviour.

\begin{question}{Exercise}\vspace{-12pt}
	\begin{prob}
		\begin{quotation}
		More precisely, the above statement is as follows: if $\vec F(t)$ 
        is a homogeneous force field and $V(t,\vec x)=
        -\vec x\cdot\vec F(t)$ the corresponding potential 
        that appears in the \schr\ equation, then 
        any solution $\psi$ of the latter is of the form 
        $\bigl(\exp(\mathrm{i}\alpha)\,\psi'\bigr)\circ\Phi^{-1}$, where 
        $\psi'$ is a solution of the free \schr\ equation,
        $\Phi:\mathbb{R}^4\rightarrow\mathbb{R}^4$
        is the flow map $(t,\vec x)\mapsto\Phi(t,\vec x):=(t,\vec x+\vec\xi(t))$, corresponding to a classical solution of 
        $m\ddot{\vec\xi}(t)=\vec F(t)$ with initial condition 
        $\vec\xi(t=0)=\vec 0$ and arbitrary initial velocity, 
        and $\alpha(t,\vec x)$ is a phase factor for which an 
        explicit integral formula can be written down 
        depending on $\vec\xi(t)$.
        
        Prove this theorem and derive the integral formula for $\alpha(t, \vec x)$. (If you get stuck, see \cite[section\,3]{giuliniEP2012}.)
		\end{quotation}
	\end{prob}
\end{question}

Finally we also emphasise that the often used 
`equality of inertial and gravitational 
mass' expresses the equivalence principle only in 
restricted physical situations, and then only 
in the context of Newtonian gravity. Only if 
the Newtonian laws of free fall apply does 
that equality ensure the universality of 
free fall (for unstructured bodies). On the 
other hand, in GR, we do not have an unambiguous 
way to even define the notion of (passive) 
gravitational mass of a body interacting with 
others. Here, the equivalence principle should 
really be formulated in an invariant way that is 
independent of representation dependent 
definitions of `masses'. We will have much 
more to say about this in section~\ref{subsec:anomalous-couplings}.

\subsection{Forces versus inertial structure}
At this point we also wish to recall another aspect 
concerning the dichotomy between our theory of 
gravity on one hand, and our theories of 
fundamental matter on the other. The latter 
comprise the standard model, which is a field-theoretic 
description of the weak, strong, and electromagnetic
forces. On the other hand, according to GR,
gravity is not a \emph{force} but rather unified with 
the inertial structure. For that reason one sometimes 
speaks of the `gravito-inertial field' (rather than just 
`gravitational field'). The geometric structure 
representing this field is the connection. Only 
\emph{after} the connection is known, and the 
inertial motion thereby specified,  does it make 
sense to speak of forces: a force is, by definition,  
the cause for deviations from inertial motion. 

A mathematically distinguishing feature of the 
gravito-inertial field in comparison with other 
fundamental fields, that is often not sufficiently 
appreciated, is that the set of connections 
is an affine space, not a vector space (like the 
set of sections in a vector bundle). This means 
that there is no such thing as a `zero connection'
and hence no spacetime `free' of gravitation/inertia. 
True, a flat connection may---in affine charts---be 
represented by identically vanishing coefficients, like
in Minkowski space endowed with affine (inertial) coordinates,
but that clearly does not mean that inertia---and hence 
gravity---is somehow `absent' in any reasonably 
invariant way. Hence, whereas it makes sense to say that 
a certain region of spacetime is `free' of electromagnetic 
fields, it makes no sense to make such a statement for 
gravity. 

In passing we note that the notion of a `geometric object'
is not at all restricted to tensor fields; see, e.g., 
\cite{salvioliGeomObject1972}. Tensor fields transform 
linearly under changes of coordinates (or diffeomorphisms)
whereas connections have an 
affine (i.e.\ inhomogeneous) transformation property. Yet 
it clearly remains true in both cases that the coefficients 
are known in \emph{all} coordinate systems if they are 
known in a single one. That, in fact, is often taken 
as a working definition of a `geometric object'; e.g.,  \cite[\S\,4.13, pp.\,84--87]{trautmanFoundGR1964}. 
In particular, this applies to connections on the tangent bundle (like the Levi--Civita connection) that can be used together with the metric to 
algebraically form other geometric objects, which need 
not be tensors. 

In order for the above statement to make sense, that 
a certain region of spacetime is `free of gravity', 
`gravity' would have to be identified with a tensor 
field with all tensor-space values, in particular 
zero, as admissible states.
Note that the metric itself does not fall into that 
class, since, e.g., the zero section is not 
admissible so that the set of gravitational states 
is not a vector space. Such a vector space structure 
is achieved if one chooses a fixed background 
connection as reference and represents
the `gravitational field' as the \emph{difference} 
of the physical connection to that reference. 
But then the latter defines a background structure 
that explicitly breaks diffeomorphism invariance 
down to the subgroup of those diffeomorphisms 
preserving that background. Alternatively, 
`zero gravity' is  also sometimes taken to 
mean `zero curvature' \cite{syngeGR}. 
But that re-introduces the old dichotomy between 
gravity and inertia (inertial forces clearly exist 
in flat Minkowski space) that GR had so 
successfully overcome, their unification being the heart of 
the equivalence principle, just like background 
independence is.\footnote{Einstein considered 
the unification of inertia and gravity to be 
\emph{the} distinctive physical 
achievement of GR, not the fact that it 
can be formulated in purely geometric terms,
which he regarded more a matter of semantics
rather than physics; 
see~\cite{lehmkuhlWhyEinsteinDid2014}.}
For further discussions of these 
general aspects we refer to 
\cite{giuliniTraegheit2002,giuliniGenCov2007,pfisterInertiaAndGravitation2015}.

The inertial structure of flat Minkowski space is 
also deeply rooted in its symmetry properties and 
hence in all theories of interactions except gravity. 
In fact, the inertial structure endows Minkowski 
space with the structure of a four-dimensional real 
affine space in which an open subset in the set of 
all `straight lines', namely the timelike ones, 
represent inertial motion of `test particles'. 
It can be shown \cite{alexandrov1975} that the subgroup 
within the group of bijections of Minkowski space 
that consists of those bijections that preserve this inertial structure 
(i.e.\ map timelike straight lines to timelike straight
lines) is the group generated by Poincar\'e 
transformations and homotheties%
\footnote{%
Homotheties are scalings about any centre point, 
i.e.\ maps of the form $x\mapsto x':= a(x-x_0)+x_0$, 
where $x_0$ is any point in Minkowski space, 
the `centre' of the homothety, and $a$ is a 
non-zero real number, its `scaling' parameter.}.
The latter are eliminated if, e.g., a measure of 
length is provided at least along any of the 
inertial straight lines (a `clock'). In that 
sense the inertial structure alone almost 
determines the Poincar\'e group without 
further assumptions. Note in particular 
that no continuity or other regularity assumption 
enters the proof of this result.  Now, according 
to Felix Klein's `Erlanger Programm' 
\cite{kleinErlangerProgramm} geometries and their 
automorphism groups are two sides of the same coin. 
Hence, the Poincar\'e group, which lies at the heart 
of fundamental theories of matter, is the algebraic 
expression of the inertial structure, which is of 
geometric nature.

In special-relativistic theories 
the inertial structure is fixed once and for all in 
a way that is entirely independent of the matter 
content. In contrast, the dynamics of the matter 
clearly \emph{does} depend on the inertial 
structure. Hence this dependence is unidirectional. 
In that sense Minkowski space and the Poincar\'e 
group are absolute structures, something to be 
abandoned according to the principles that led to 
GR.  An alternative approach how to give up the absolute structure
of Minkowski space is to `gauge' Poincar\'e symmetry, as discussed
in this volume's Chapters on Poincar\'e gauge gravity and teleparallel gravity.  That 
abandoning global Poincar\'e symmetry will likely imply 
a major revision in the concepts of \QFT\ should 
be obvious by once more recalling 
the central role the Poincar\'e group and its 
representation theory plays there, e.g., regarding 
the concepts of `particle' and `elementary'.

For the highly idealised concept of test particles, 
an inertial structure reduces to the concept of a 
\emph{path structure}. 
The latter assigns a unique path in spacetime to 
any pair consisting of a point and one-dimensional 
subspace in the tangent space at that point. 
In other words, the path is universal for all
`test particles', only depending on the 
initial conditions, but not on other contingent 
properties the particle may still have. The path 
structure in GR is special in a 
twofold way: first, the paths are geodesics for 
\emph{some} linear connection; second, the connection 
is the Levi--Civita connection for \emph{some} 
metric. It has been worked out in precise 
mathematical terms how to characterise such 
special path structures
\cite{ehlers-koehler1977,coleman-korte1980}. 

The important point we wish to stress 
and keep in mind at this point is the 
distinction between `forces' on one 
side, and `inertial structure' on the 
other. The fields in the standard model 
account for forces, whereas the field in 
GR accounts for the inertial 
structure. The former have a natural `zero'
value, representing physical absence, but that 
does not apply for the latter; spacetime cannot 
be without inertial structure. Moreover, the 
former are quantised, the latter 
is---so far---only understood classically.  

Different attitudes exist as to whether and how 
the conceptual difference just discussed should be 
overcome. Whereas some feel it would be desirable to 
reformulate gravity in a way more like the 
other `forces', others believe just the opposite and 
stress the physical significance of that 
difference. For example, in the first case, 
the `graviton'---a word often epitomising the
believed quantum nature of gravity---is taken to 
mediate gravity in the same sense as the other 
gauge bosons mediate the other forces.  
In the second case, this would not make sense and 
the programme to `quantise gravity' is presumably 
no more plausible than that of `general-relativising' 
RQFT or `gravitizing Quantum Mechanics' \cite{penroseGravitizationQM2014}.

Another attitude towards the relation between 
\emph{gravity and matter} or 
\emph{inertial structures and forces} is that gravity 
is subordinate to matter, an accompanying phenomenon 
based on certain collective matter degrees of freedom.
This is also sometimes expressed by saying that gravity 
is an \emph{emergent} phenomenon. In this instance, GR would
have a status comparable to, say, the Navier--Stokes 
equation of hydrodynamics, which also describes 
collective degrees of freedom that eventually can be 
reduced to those of the fluid's constituents. In such 
a picture, the microscopic state of matter fully 
determines the state of the gravitational field, which 
has no own propagating degrees of freedom. Gravitational 
waves are then, like sound waves, collective excitations
of an underlying and more fundamental field of matter.  
The idea that the state of matter and their interactions 
should completely determine the inertial structure 
predates GR and goes back to the critique that Ernst Mach 
voiced in his book \cite{machMechanik1883} on mechanics 
on Newton's interpretation of his [Newton's] famous 
bucket experiment. Einstein made this idea into what he 
called `Mach's Principle' on which he based many 
heuristic considerations during the formative years of 
GR. Remarkably, even in 1918, well more than two years after 
the final formulation of GR, Einstein explicitly named 
Mach's Principle as an essential and indispensable part of GR,
next to the principle of relativity and the principle of 
equivalence \cite{einsteinPrinzipiellesART1918}.%
\footnote{For a comprehensive account on the meanings 
and significance of Mach's Principle see \cite{barbour-pfisterMachsPrinciple}.}   

Clearly, the mathematical structure of GR does not 
support Einstein's version of Mach's principle. 
According to GR, the gravitational field has its own 
degrees of freedom that propagate causally, albeit the 
causal structure with respect to which this statement 
is true is not a fixed background structure but rather 
determined by the evolving field itself. That remark 
holds irrespective of gauge dependent appearances of 
Einstein's equations, in which certain components of the 
field may seem to propagate instantaneously due to 
the fact that they obey elliptic rather than hyperbolic 
equations. But that is just as deceptive as in ordinary 
electrodynamics, where, e.g., in the Coulomb gauge the 
scalar potential obeys an elliptic Poisson equation 
whereas the transversal part of the vector potential 
obeys a hyperbolic d'Alembert equation with respect to
the transverse current density\footnote{Note that the 
transverse current density is a non-local and non-causal 
function of the physical current density.} as source;
see, e.g., \cite{brillCausalityCoulomb1967} and 
\cite[exercise 6.20, pp.\,291--292]{jacksonED3rd}.
This remark becomes important in the recent debates on 
alleged inferences of quantisation of gravity from gravitationally 
induced entanglement; see \cite{fragkosInference2022} and 
references therein.

\subsection{Foundations of General Relativity and the role of Quantum Mechanics}
One way to `understand' the foundations of a 
physical theory is to axiomatise it, that is, 
to rigorously deduce it from a minimal set of 
assumptions. The latter should be operationally 
meaningful though they will in most cases be highly 
idealised. Attempts to axiomatise GR 
face the problem to somehow represent `clocks' 
and `rods', which from a physical point of view 
are highly complex systems. The strongest simplification 
is to reduce them to point particles and light rays, 
by means of which one may define an inertial structure 
and a conformal structure as in the EPS scheme of 
Ehlers, Pirani, and Schild \cite{EPS1972,EPS2012}.
In that scheme the `particles' represent the 
inertial structure and are given by unparameterised 
timelike worldlines, which the axioms force to be 
unparameterised geodesics (also 
known as autoparallels) of a torsion-free 
connection. The `particles' are therefore just an 
equivalence class of torsion-free connections which 
share the same autoparallels.%
\footnote{Two connections 
$\Gamma^\lambda_{\mu\nu}$ and 
$\hat\Gamma^\lambda_{\mu\nu}$ are equivalent in that 
sense if and only if there exists a covector field 
$V_\mu$ such that $\hat\Gamma^\lambda_{\mu\nu}-
\Gamma^\lambda_{\mu\nu}=\delta^\lambda_\mu V_\nu-
\delta^\lambda_\nu V_\mu$.}
The `light rays' in the EPS scheme
determine a conformal equivalence class 
of metrics. A compatibility axiom requires that the 
`light rays' are suitable limits of the 
`particles', i.e.\ that the generators of the metric 
light cones (which are conformal invariants) are 
suitable limits of the unparameterised timelike 
geodesics. All this does not yet lead to a 
semi-Riemannian structure but rather to a Weyl structure
consisting of a triple $(\mathcal{M},[g],\nabla)$, where 
$\mathcal{M}$ is a smooth 4-d differentiable manifold, $[g]$
is a conformal equivalence class of metrics, and 
$\nabla$ is a torsion-free connection that satisfies 
$\nabla_\lambda g_{\mu\nu}=\varphi_\lambda g_{\mu\nu}$
for some 1-form $\varphi$ depending on the  
representative $g$ of $[g]$. If 
$g'_{\mu\nu}=\exp(\Omega)g_{\mu\nu}$ then 
$\varphi'_\lambda=\varphi_\lambda+\partial_\lambda\Omega$.
Hence, we may also identify a Weyl structure with 
equivalence classes $[(g,\varphi)]$ of pairs 
$(g,\varphi)$, where $(g,\varphi)\sim (g',\varphi')$ if and 
only if there exists a smooth function $\Omega:\mathcal{M}\rightarrow\mathbb{R}$ such that 
$g'=\exp(\Omega)g$ and $\varphi'-\varphi=d\Omega$. Such an
equivalence class $[(g,\varphi)]$ determines a unique 
torsion-free connection $\nabla$ such that for 
any representative $(g,\varphi)$ one has 
$\nabla g=\varphi\otimes g$. This would be equivalent 
to a semi-Riemannian structure if and only if $\varphi$
is exact, $\varphi=df$, in which case there is a 
representative $(g,\varphi)$ so that $\varphi=0$
and $\nabla g=0$, i.e., $\nabla$ is the Levi--Civita
connection for $g$. EPS close this gap in physical 
terms by simply stating as an additional axiom that 
there are no so-called `second clock effects', 
which amounts to just $d\varphi=0$ and hence (at 
least local) exactness. 

Now, a somewhat surprising result is that instead of this postulate, the gap can
also be closed by the requirement that the WKB limit of the
massive Dirac and/or Klein--Gordon field in a Weylian 
spacetime is such that the rays (integral lines of 
the gradient of the eikonal) are just the `particles'
\cite{audretschWeylRiemann1983}. This is an entirely 
new aspect concerning the relation between QM and GR. 
It suggests that QM can, in fact, also play a positive 
role in laying the foundations to GR, rather than just 
cause trouble. An unexpected twist to the story indeed!

%%%%%%%%%%%%%%%%%%%%%%%%%%%%%%%%%%%%%%%%%%%%%%%%%%%%%%%%%%%%%%%%%%%%%%
%%%%%%%%%%%%%%%%%%%%%%%%%%%%%%%%%%%%%%%%%%%%%%%%%%%%%%%%%%%%%%%%%%%%%%

\section{Quantum matter on classical gravitational backgrounds}
\label{sec:qm-class-backgrounds}

In this section, we will discuss the description of of quantum matter systems on a fixed classical background spacetime---i.e.\ we consider the matter fields as `test systems' probing the gravitational field and neglect their backreaction onto it. As explained at the end of section \ref{subsec:whyCare}, the generally accepted theoretical framework for this setting is QFTCS. However, notwithstanding its formal and conceptual successes, QFTCS is mathematically and conceptually very `heavy', and its relation to (approximately) Galilei-symmetric
physics, which for example is relevant for the description of quantum-optical experiments (as discussed in this volume's Chapter on quantum tests of gravity), is not as well understood as one might suspect.

As long as one is only interested in the effect of background \emph{Newtonian} gravity on quantum systems (the experimental study of which is discussed in the Chapter on quantum tests of gravity as well),
this does not pose a problem: one may simply include the background Newtonian gravitational potential $\Phi$ into the Hamiltonian describing the system in the usual way, leading to a \schr\ equation that for a single bosonic particle of mass $m$, only subject to gravity, is of the form
\begin{equation}
    \I\hbar\partial_t \Psi = \left(-\frac{\hbar^2}{2m}\laplacian + m \Phi\right) \Psi.
\end{equation}
This simple coupling of Galilei-invariant QM to Newtonian gravity has been extensively tested in the gravitational field of the earth, beginning with neutron interferometry in the Colella--Overhauser--Werner (COW) experiment in 1975 \cite{colellaObservationGravitationallyInduced1975} and leading to modern light-pulse atom interferometers, which for example provide the most sensitive gravimeters to date \cite{karcherImprovingAccuracyAtomInterferometers2018}. However, as soon as one is interested in \emph{post}-Newtonian effects of gravity on quantum systems, one needs a method either to describe the respective situation in terms of QFTCS proper or to include `post-Newtonian corrections' into the \schr\ equation above. Note that such post-Newtonian effects can be of two different kinds: they either correspond to post-Newtonian couplings of the `Newtonian' potential $\Phi$, or are couplings to those parts of the multi-component Einsteinian gravitational field that---as a matter of principle---simply do not exist in the scalar Newtonian theory, i.e.\ couplings to the vectorial (gravitomagnetic) and tensorial (gravitational waves) components.

Apart from the obvious fundamental theoretical interest in post-Newtonian gravitational effects in quantum systems, recently there has also been increased interest from an experimentally-oriented point of view, since the ever-increasing precision of quantum
experiments is expected to allow for the detection of novel `relativistic effects' that were not considered before. In particular, for composite systems one expects post-Newtonian couplings between internal and external (i.e.\ centre-of-mass) degrees of freedom of both non-gravitational and gravitational origin, which might lead, e.g., to quantum dephasing \cite{zychQuantumInterferometricVisibility2011,pikovskiUniversalDecoherence2015,bonderGravityEmergenceClassicality2015}.

The occurrence of such post-Newtonian couplings may on a heuristic level be understood based on the notion of `relativistic effects' known from classical physics: formally replacing, in the spirit of the `mass defect', the mass parameter in the single-particle Hamiltonian for a quantum system under Newtonian gravity by the rest mass $M$ of a composite system plus its internal energy $H_\text{int}$ divided by $c^2$, we obtain a coupling between the internal and external degrees of freedom according to
\begin{subequations} \label{eq:coupling_int_ext_heuristic}
\begin{align}
    H_\text{single particle}(\vec p,\vec r,m) &= \frac{\vec p^2}{2m} + m \Phi(\vec r) \\
    \to H_\text{composite} &= H_\text{single particle}(\vec P,\vec R,M + H_\text{int}/c^2) \nnl
        &= \frac{\vec P^2}{2(M + H_\text{int}/c^2)} + (M + H_\text{int}/c^2) \Phi(\vec R) \nnl
        &= \frac{\vec P^2}{2M}\left(1 - \frac{H_\text{int}}{Mc^2}\right) + (M + H_\text{int}/c^2) \Phi(\vec R) + \order{c^{-4}},
\end{align}
\end{subequations}
where $\vec P$ and $\vec R$ denote the `central coordinates' of the composite system, i.e.\ total momentum and some appropriate centre of mass coordinate. Another way to heuristically motivate such a post-Newtonian internal--external coupling is the replacement, in the \schr\ equation describing the internal dynamics, `$t \to \tau$' of coordinate time by proper time of the central worldline---since the internal degrees of freedom experience special-relativistic and gravitational time dilation, so the argument goes, they will become coupled to the central dynamics \cite{zychQuantumInterferometricVisibility2011,pikovskiUniversalDecoherence2015,lorianiInterferenceClocks2019}.

Of course, such heuristic motivations of post-Newtonian couplings have great suggestive value. They are, however, conceptually dangerous for several reasons: Firstly, such treatments guarantee neither completeness nor independence of the suggested `novel relativistic effects'---one might, on the one hand, overlook some effects, while on the other hand double-counting others (as would have been the case in our example above, had we included \emph{both} the `mass defect' and `relativistic time dilation'). Secondly, those descriptions rely fundamentally on semiclassical notions such as central worldlines, and thus presuppose (a) separability (at least approximately) of the total state of the system into an external and an internal part---even though interactions, leading to entanglement, are precisely the point of interest!---, and (b) a semiclassical nature of the external state. Of course, for some applications---e.g.\ in atom interferometry---these assumptions may be well-suited; nevertheless a fundamental understanding of the post-Newtonian coupling of quantum systems to gravitational fields should be \emph{independent} of such assumptions.

Therefore, to correctly and completely describe quantum systems in gravitational fields, a systematic treatment is needed, starting from well-established first principles and properly \emph{deriving} a full theoretical description. Only such a systematic approach is \emph{complete} and \emph{free of redundancies}, and therefore can allow for reliable predictions when applied, for example, to experimental situations. As said before, in the end such a properly relativistic description should emerge from QFTCS. However, as long as we are only interested in low-order `relativistic corrections' to the Newtonian coupling to gravity, consider approximately stationary spacetimes (such that there is a consistent notion of particles), and stay below the energy threshold of pair production, we may avoid the use of the complex framework of QFTCS. Instead, we may restrict ourselves to a fixed-particle sector, in which the theory is effectively described by classical field equations, and in this sector employ a post-Newtonian expansion framework as an easier method for systematic description of quantum systems under gravity. This essentially seems to be the only straightforward way to apply the equivalence principle to usual Galilei-invariant QM: to first deform Galilei to Poincar\'e invariance (the deformation parameter being $c^{-1}$), then apply the minimal coupling scheme, dictated by the equivalence principle and providing the couplings to the gravitational field at least up to additional curvature terms, and then contract again by a post-Newtonian expansion in the deformation parameter $c^{-1}$. 

In section \ref{subsec:description-model-systems}, we are going to describe such a systematic post-Newtonian expansion framework applicable to the description of model quantum systems. This framework enables us in particular to derive a full description of a simple toy model of a two-particle atom to order $c^{-2}$, which will be explained in section \ref{subsubsec:composite-systems}. In particular, this provides a proper derivation of the above-mentioned internal--external couplings. However, we will encounter `anomalous' couplings of the internal kinetic and potential energies to the Newtonian gravitational potential: depending on the choice of coordinates in which the internal degrees of freedom are described, one does not get the heuristically expected coupling from \eqref{eq:coupling_int_ext_heuristic}, but the different energy forms couple differently. This phenomenon of `anomalous' couplings appears quite generally in the context of the post-Newtonian description of composite systems under gravity. In section \ref{subsec:anomalous-couplings}, we will discuss these issues from a more general point of view, in the context of diffeomorphism-invariant field theory: we will make precise and critically evaluate arguments by Carlip~\cite{carlipKineticEnergyEquivalencePrinciple1998} about the emergence of such `anomalous' couplings and their possible elimination by application of 
diffeomorphisms. The content of 
sections~\ref{subsec:description-model-systems} already appeared 
in the cited literature, whereas that of 
section~\ref{subsec:anomalous-couplings} is new.

\subsection{Systematic description of model systems in post-Newtonian gravity}
\label{subsec:description-model-systems}
In order to be able to perform a post-Newtonian expansion of a (locally) Poincar\'e-invariant theory, which perturbatively includes post-Newtonian effects on top of a Newtonian description, we need some \emph{background structures} that enable us to speak of `weak gravitational fields', `small velocities', and space and time as separate notions. For these background structures, we take a \emph{background metric} in combination with a \emph{background time evolution vector field}, which is hypersurface orthogonal, timelike, and of constant length with respect to the background metric: The time evolution field gives us a $3+1$ decomposition of spacetime into time (its integral curves) and space (the leaves of its orthogonal distribution), while also being a reference for the definition of `slow movements'. Similarly, the background metric serves as reference point for the `absence' of gravitational fields.\footnote
    {The size of deviations from these reference points is measured by means of the Euclidean metric defined by the (Lorentzian) background metric $g^{(0)}$ and the time evolution vector field $u$ as $g_\text{Euc} := g^{(0)} - 2 \frac{u^\flat \otimes u^\flat}{g^{(0)}(u,u)}$, where $u^\flat = g^{(0)}(u,\cdot)$ is the one-form associated to $u$ via $g^{(0)}$.}

Concretely, we take as background spacetime \emph{Minkowski spacetime} $(\mathcal M, \eta)$ and as background time evolution vector field a timelike geodesic vector field $u$ on $(\mathcal M, \eta)$, with Minkowski square $\eta(u,u) = -c^2$ (where $c$ is the velocity of light). $u$ can be interpreted as the four-velocity field of a family of inertial observers in background Minkowski spacetime to which we will refer our post-Newtonian expansions. The \emph{physical} spacetime metric $g$ will be a perturbation on top of the background Minkowski metric $\eta$. On the three-dimensional leaves of `space', which are $\eta$-orthogonal to $u$, we have two Riemannian metrics: a flat metric $\delta$ induced by the background metric $\eta$, as well as the physical spatial metric $\gspac$ induced by the physical metric $g$.

Introducing coordinates $(x^\mu) = (x^0 = ct, x^a)$ adapted to the background structures, i.e.\ such that in these coordinates the Minkowski metric takes its usual component form $(\eta_{\mu\nu}) = \mathrm{diag}(-1,1,1,1)$ and the time evolution field is $u = \partial/\partial t$, we may express all the following post-Newtonian computations in those coordinates, which will make the results look like usual Newtonian results plus post-Newtonian corrections. Note, however, that the results themselves are of course independent of the coordinates chosen to express them.

As the expansion parameter in powers of which we organise our post-Newtonian expansion, we take the inverse of the velocity of light, $c^{-1}$. That is, we will expand all relevant quantities as formal power series in the parameter $c^{-1}$. The term of order $c^0$ in such a series corresponds to the Newtonian limit, while the higher-order terms are the post-Newtonian corrections. Formally, the Newtonian limit is obtained as the $c\to\infty$ limit of the series. Of course, analytically speaking, a `Taylor expansion' in a dimensionful parameter like $c$ does not make sense (and even less so any limit in which $c$ is varied, since it is a constant of nature); only for dimensionless parameters can a meaningful `small-parameter approximation' be made. Nevertheless, as a convenient device to keep track of post-Newtonian effects, such a formal expansion is perfectly fine, enabling us to view the post-Newtonian theory as a formal deformation of its Newtonian counterpart.\footnote
    {In physical realisations of a Newtonian limit, the corresponding expansion parameter has to be chosen as the dimensionless ratio of some typical velocity of the system under consideration to the speed of light. For some discussion of the relationship of formal `$c\to\infty$' limits to actual physical approximations, see, e.g., section II B of reference \cite{tichyCovariantFormulationPost1Newtonian2011}.}
Note also that for the expansion of some objects, terms of negative order in $c^{-1}$ will appear, such that strictly speaking we are dealing with formal Laurent series. For a series with non-vanishing terms of negative order, the formal Newtonian $c\to\infty$ limit does not exist.

In order to obtain a consistent Newtonian limit, we need to identify the time coordinate $t$ (defined as the flow parameter along the background time evolution field $u$) with Newtonian absolute time. Therefore, we have to treat $t$ as being of order $c^0$ in our formal post-Newtonian expansion. Note that this implies that the timelike coordinate $x^0 = ct$ with physical dimension of length is to be treated as of order $c^1$, differently to the spacelike coordinate functions $x^a$ to which we assign order $c^0$. This necessity of treating the time direction differently from spacelike directions in the consideration of the formal relationship between Poincar\'e- and Galilei-symmetric theories is well-known: it arises, for example, in the context of Newton--Cartan gravity (i.e.\ geometrised Newtonian gravity) \cite{ehlersUeberNewtonschenGrenzwert1981,ehlersNewtonianLimit2019}, or in the Lie algebra contraction from the Poincar\'e to the Galilei algebra \cite{inonuContractionGroups1953,bacryPossibleKinematics1968}. Due to this, for example the Minkowski metric
\begin{equation}
    \eta = \eta_{\mu\nu} \, \D x^\mu \otimes \D x^\nu = - c^2 \D t^2 + \D \vec x^2
\end{equation}
consists of a temporal part of order $c^2$ and a spatial part of order $c^0$.

We expand the components of the inverse of the physical spacetime metric as formal power series
\begin{equation} \label{eq:exp_metric}
    g^{\mu\nu} = \eta^{\mu\nu} + \sum_{k=1}^\infty c^{-k} g^{\mu\nu}_{(k)} \; ,
\end{equation}
with the lowest-order terms assumed to be given by the components of the inverse Minkowski metric, and the higher-order coefficients $g^{\mu\nu}_{(k)}$ being arbitrary. Likewise, this could have been given by the corresponding power series expansion of the metric; here we choose to expand the inverse metric for later notational simplicity.

An important example for a post-Newtonian metric is the \emph{Eddington--Robertson parametrised post-Newtonian metric}, 
given by
\begin{equation} \label{eq:ER_PPN_metric}
    g^\text{ER--PPN} 
    = \left(-1 - 2 \frac{\Phi}{c^2} - 2 \beta \frac{\Phi^2}{c^4}\right) c^2 \D t^2 
        + \left(1 - 2 \gamma \frac{\Phi}{c^2}\right) \D \vec x^2 
        + \order{c^{-4}}.
\end{equation}
Note that this is the power series expansion of the metric itself and not its inverse.
The dimensionless \emph{Eddington--Robertson parameters} $\beta$ and $\gamma$ account for possible deviations from GR.
For the case of GR, which corresponds to the values $\beta = \gamma = 1$,
the metric \eqref{eq:ER_PPN_metric} solves the Einstein field equations in a $c^{-1}$-expansion for a static source, with $\Phi$ being the Newtonian gravitational potential of the source. The family of metrics for different values of $\beta,\gamma$ then form a `test theory', enabling tests of GR against possible different metric theories of gravity. The Eddington--Robertson metric is the most simple example of a metric in the general 
parameterised post-Newtonian (henceforth 
abbreviated by PPN) formalism,
which provides a general framework of metric test theories of gravity in the weak-field regime. For a detailed discussion, see \cite{willTheoryExperimentGravitationalPhysics2018,poissonWillGravity2014}.

\subsubsection{Single particles}
\label{subsubsec:single-particles}

Based on the post-Newtonian expansion framework introduced above, we now discuss the systematic description of single, free quantum particles in post-Newtonian gravitational fields. Performing a WKB-inspired formal expansion of classical relativistic field equations, we will arrive at a \schr\ equation with post-Newtonian corrections \cite{kieferQuantumGravitational1991,laemmerzahlHamiltonOperator1995,giuliniSchrodingerNewtonNonrelativisticLimit2012,schwartzPostNewtonianCorrections2019,schwartzPostNewtonianDescription2020}. We will motivate the consideration of the classical field equations from QFTCS, and briefly explain the expansion of the minimally coupled Klein--Gordon equation as an example. Details of the discussion may be found in~\cite{schwartzPostNewtonianCorrections2019,schwartzPostNewtonianDescription2020}. Related discussions for a massive Dirac field
are contained in \cite{ito2021,perche2021}, and 
in terms of a systematic post-Newtonian approximation in Fermi
normal coordinates with respect to a rotating frame along an
accelerated worldline in \cite{alibabaei2022}.

The consideration of the \emph{classical} relativistic field theories for the description of the one-particle sector of the \emph{quantised} theory may be motivated as follows. On a globally hyperbolic, stationary spacetime, for a given relativistic free field theory, there is a preferred Fock space representation of the field observables in QFTCS (and thus a preferred notion of particles). Loosely speaking\footnote
    {For details and caveats of the construction, we refer to the extensive discussion in the monograph by Wald~\cite{waldQuantumFieldTheory1994}; see section~4.3 and references therein for the case of Klein--Gordon fields, as well as section~4.7 for fermionic and other higher-spin fields.},
the construction of this representation works as follows \cite{waldQuantumFieldTheory1994}: We consider classical solutions of the given relativistic field equation, and among those we pick out the subspace of `positive-frequency solutions' with respect to the stationarity Killing field. Completing the positive-frequency subspace with respect to the `field inner product' (i.e.\ the Klein--Gordon inner product for Klein--Gordon fields, or the Dirac inner product for Dirac fields, etc.), we obtain a `one-particle Hilbert space'. The Hilbert space for the \QFT, on which the field operators can be represented, is then the bosonic or fermionic Fock space over this `one-particle space', according to the spin of the field.

This means that, in the framework of QFTCS, the one-particle sector of a given free relativistic quantum field theory on a globally hyperbolic stationary spacetime is described by the positive-frequency solutions (in an appropriate sense) of the classical field equation, with the corresponding inner product. This is the underlying reason for the effective description of the one-particle sector of the quantum theory by the \emph{classical} field theory---which in the literature is often called consideration of the `first-quantised theory'---working so well, as long as one is concerned only with processes far enough below the threshold of pair production.

For a non-stationary spacetime, the above motivation of course breaks down: there is no time translation symmetry and, therefore, not even a natural notion of particles in QFTCS. However, as long as we assume the background time evolution vector field $u$ to be \emph{approximately} Killing for the physical spacetime metric $g$, on a heuristic level we can still expect perturbative `positive-frequency' solutions of classical field equations to lead to approximately correct predictions regarding the observations of observers moving along the orbits of $u$. Thus, from QFTCS we are led to the consideration of formal perturbative expansions of relativistic field equations on curved spacetimes as the natural means for the description of single quantum particles in post-Newtonian gravitational fields.

Assuming a formal power series expansion of the inverse physical spacetime metric as in \eqref{eq:exp_metric}, we are able to (in principle) obtain the post-Newtonian expansion of any relativistic field equation in complete generality, by making a WKB-inspired power series ansatz for positive-frequency solutions. Using this procedure, we may obtain the \schr\ equation that describes the corresponding one-particle states to arbitrary post-Newtonian order, in a completely systematic fashion. Specifically, for a minimally coupled scalar field, a brief outline of this procedure is as follows (see~\cite{schwartzPostNewtonianCorrections2019, schwartzPostNewtonianDescription2020} for details):

The minimally coupled Klein--Gordon equation is
\begin{equation} \label{eq:KG}
    \left(\Box - \frac{m^2 c^2}{\hbar^2}\right)\Psi_\mathrm{KG} = 0,
\end{equation}
where $\Box = g^{\mu\nu} \nabla_\mu \nabla_\nu$ is the d'Alembert operator of the spacetime metric $g$. Expressing the covariant derivatives in terms of the components of the metric and inserting the formal power series expansion \eqref{eq:exp_metric}, we obtain an expansion of the d'Alembert operator in powers of $c^{-1}$. In this process, we have to express the expansion coefficients of the \emph{metric} components $g_{\mu\nu}$ in terms of those of the \emph{inverse} metric components $g^{\mu\nu}$, which is possible by use of a formal Neumann series and the Cauchy product formula for products of infinite series. We then make the WKB-inspired ansatz
\begin{equation} \label{eq:WKB_ansatz}
    \Psi_\mathrm{KG} = \exp\left(\frac{\I c^2}{\hbar}S\right) \psi \,,
    \qquad
    \psi = \sum_{k=0}^\infty c^{-k} a_k
\end{equation}
for the Klein--Gordon field \cite{giuliniSchrodingerNewtonNonrelativisticLimit2012}, where we assume $S, a_k$ to be independent of the expansion parameter $c^{-1}$. Inserting this ansatz into the Klein--Gordon equation \eqref{eq:KG} with expanded d'Alembert operator, we may then compare coefficients of powers of $c^{-1}$ in order to obtain equations for $S$ and the $a_k$. At the lowest ocurring order $c^4$, we obtain that $S$ depends solely on time; the equation at order $c^3$ is then identically satisfied. At order $c^2$, the equation gives us $(\partial_t S)^2 = m^2$; since we are interested in positive-frequency solutions, we choose
\begin{equation}
    S = -mt
\end{equation}
(discarding the constant of integration, which would lead to an irrelevant global phase). 
This means that for such solutions, the function $\psi$ from \eqref{eq:WKB_ansatz} is the Klein--Gordon field with the `rest-energy phase factor' $\exp(-\I m c^2 t/\hbar)$ separated off. At order $c^1$, the Klein--Gordon equation then leads to the requirement
\begin{equation}
    g_{(1)}^{00} = 0
\end{equation}
for the metric, which is satisfied in any metric theory of gravity that reproduces the correct equations of motion for test particles in the Newtonian limit.

\begin{question}{Exercise}\vspace{-12pt}
	\begin{prob}
		\begin{quotation}
		\begin{enumerate}
\item[\textbf{(i)}] Compute the derivatives $\partial_\mu \Psi_\mathrm{KG}$ and
  $\partial_\mu\partial_\nu \Psi_\mathrm{KG}$ of the ansatz
  \eqref{eq:WKB_ansatz} for the Klein--Gordon field.

\item[\textbf{(ii)}] Using the derivatives of $\Psi_\mathrm{KG}$ and the d'Alembert
  operator $\Box = - c^{-2} \partial_t^2 + \Delta$ in Minkowski
  spacetime, determine the lowest-order terms in the $c^{-1}$
  expansion of the Klein--Gordon equation \eqref{eq:KG}.  Show that
  the leading-order equation, at order $c^4$, leads to $\partial_t S =
  0$.  Show that the equation at order $c^3$ is identically satisfied,
  and use the order $c^2$ equation to determine $S$.  Finally, show
  that the $c^0$ term leads to the free Schrödinger equation for
  $a_0$.

\item[\textbf{(iii)}] Instead of Minkowski spacetime, we now consider a `Newtonian'
  metric given by $g = - (1 + 2 \frac{\phi}{c^2}) c^2 \D t^2 + \D\vec
  x^2 + \order{c^{-2}}$.  For a general metric $g$, the d'Alembert
  operator acting on functions is given by
  \begin{equation*}
    \Box f = \nabla^\mu \nabla_\mu f
    = \frac{1}{\sqrt{-g}} \partial_\mu (\sqrt{-g}
      g^{\mu\nu} \partial_\nu f),
  \end{equation*}
  where $\sqrt{-g}$ denotes the square root of minus the determinant
  of the matrix of metric components.  Compute $\Box f$ for the
  Newtonian metric to order $c^{-2}$ in terms of partial space and
  time derivatives $\partial_a, \partial_t$.  (Be careful: $\partial_0
  = \frac{\partial}{\partial(ct)} = c^{-1} \partial_t$!)

  Analogously to the Minkowski calculation, show that the $c^{-1}$
  expansion of the Klein--Gordon equation in the Newtonian metric to
  order $c^0$ leads to a \schr\ equation for $a_0$ including the
  Newtonian potential $\phi$.

  \emph{Hint: the inverse metric has components $g^{00} = - 1 + 2
    \frac{\phi}{c^2} + \order{c^{-4}}$, $g^{0a} = \order{c^{-3}}$, $g^{ab} =
    \delta^{ab} + \order{c^{-2}}$.  For the determinant term, use
    $\frac{1}{\sqrt{-g}} \partial_\mu \sqrt{-g} = - \frac{1}{2}
    g_{\rho\sigma} \partial_\mu g^{\rho\sigma}$.}
\end{enumerate}
		\end{quotation}
	\end{prob}
\end{question}

Using these results, we can then obtain a general, expanded version of the Klein--Gordon equation. This fully expanded equation is rather horrendously complicated (see the appendix of \cite{schwartzPostNewtonianCorrections2019}), but it can nevertheless be written down in its entirety. From it, we may read off, order by order, equations for the coefficient functions $a_k$. After obtaining the equations for all the $a_k$ up to a fixed order $n$,\footnote
    {In this process, when considering higher orders, the equations for $a_k$ begin to involve time derivatives of the lower-order functions $a_l$. To eliminate those, we have to re-use the already derived equations for the lower-order $a_l$, in order to in the end obtain an equation for $\psi$ which takes the form of a \schr\ equation.}
we may recombine them into an equation for $\psi$ to order $c^{-n}$, which takes the form of a \schr\ equation plus higher-order post-Newtonian corrections. The inner product on the Hilbert space in which the `wave functions' $\psi$ described by this \schr\ equation live can be obtained by inserting the ansatz \eqref{eq:WKB_ansatz} into the Klein--Gordon inner product and expanding it to the desired order in $c^{-1}$. Since the Klein--Gordon inner product involves time derivatives of the fields, we have to use the derived \schr\ equation in this process. After this has been done, we have a concrete representation of the (approximate) one-particle sector of the Klein--Gordon quantum field theory in our post-Newtonian spacetime in terms of a \schr\ equation and an inner product for `wave functions' $\psi$ living on three-dimensional space as defined by our background structures.
Regarding the interpretation of the natural position operator that we obtain for the post-Newtonian theory, which in this representation of the Hilbert space is given by multiplication of wave functions by coordinate position $x^a$, we note that it arises---up to higher-order corrections---from the operator in the one-particle sector of Klein--Gordon theory which multiplies the Klein--Gordon fields by coordinate position $x^a$. In the case of Minkowski spacetime, this operator is the well-known Newton--Wigner position operator~\cite{newtonWignerLocalizedStates1949,schwartzClassicalNewtonWigner2020}.

As a last step, one may perform a unitary transformation such that, in the new representation, the position operator is still given by multiplication with $x^a$, but the inner product of the theory takes the form of a `flat' $\mathrm{L}^2$ inner product
\begin{equation}
    \langle \psi_\mathrm{f}, \varphi_\mathrm{f} \rangle_\mathrm{f} := \int \D^3 x \, \overline{\psi_\mathrm{f}} \, \varphi_\mathrm{f}
\end{equation}
over coordinate space $\{x^a\}$ (instead of the expanded Klein--Gordon inner product) \cite{laemmerzahlHamiltonOperator1995,schwartzPostNewtonianCorrections2019}. This means that the `flat wave functions' $\psi_\mathrm{f}$ in this representation are in fact scalar \emph{densities} on three-space. The necessary transformation may be read off order by order from the expanded Klein--Gordon inner product, and may then be used to compute the \schr\ equation in this `flat' representation.

Concretely, to first order in $c^{-1}$ the general \schr\ equation thus obtained, in the `flat' representation, takes the form
\begin{align}
    \I\hbar \partial_t \psi_\mathrm{f} 
    &= \bigg[-\frac{\hbar^2}{2m} \laplacian 
        - \frac{1}{2} \left\{g^{0a}_{(1)}, -\I\hbar\partial_a\right\} 
        + \frac{m}{2} g^{00}_{(2)} 
        + c^{-1} \bigg(\frac{1}{2m} (-\I\hbar) \partial_a \Big(g^{ab}_{(1)} (-\I\hbar)\partial_b \Big) \nnl
        &\quad + \frac{m}{2} g^{00}_{(3)} 
        - \frac{1}{2} \left\{g^{0a}_{(2)}, -\I\hbar\partial_a\right\} 
        - \frac{\hbar^2}{8m} [\laplacian \mathrm{tr}(\eta g^{-1}_{(1)})]\bigg) 
        + \order{c^{-2}}\bigg] \psi_\mathrm{f},
\end{align}
where $\laplacian = \delta^{ab} \partial_a \partial_b$ is the `flat' background Laplacian operator, and $\{A,B\} = AB + BA$ denotes the anticommutator.
Comparing this result to the $c^{-1}$-expanded classical Hamiltonian of a point particle in a curved metric, it turns out that (adopting a specific ordering scheme) all terms apart from the last one may be obtained by naive canonical quantisation of that classical Hamiltonian.

For concrete post-Newtonian metrics, some of the generic expansion coefficients from \eqref{eq:exp_metric} vanish, which simplifies the necessary computations. For the case of the Eddington--Robertson PPN metric \eqref{eq:ER_PPN_metric}, the derivation is not difficult (but somewhat tedious) to carry out completely, resulting in the Hamiltonian
\begin{subequations} \label{eq:Hamiltonian_ER}
\begin{align} \label{eq:Hamiltonian_ER_WKB}
    H^\text{ER--PPN}_\mathrm{f} 
    &= -\frac{\hbar^2}{2m} \laplacian + m\Phi
        + \frac{1}{c^2} \bigg(-\frac{\hbar^4}{8m^3} \nabla^4 
            - \frac{3\hbar^2}{4m} \gamma (\laplacian\Phi) 
            - \frac{\hbar^2}{2m} (2\gamma + 1) \Phi \laplacian \nnl
            &\qquad - \frac{\hbar^2}{2m} (2\gamma + 1) (\partial_a \Phi) \delta^{ab} \partial_b 
            + (2\beta - 1) \frac{m\Phi^2}{2} \bigg) + \order{c^{-4}}
\end{align}
in the `flat' representation.\footnote{We note that this 
Hamiltonian may also be obtained by a very similar, but 
more explicitly WKB-like method in which the logarithm 
of the Klein--Gordon field is expanded in $c^{-1}$; 
see \cite{laemmerzahlHamiltonOperator1995}.}
Comparing this to the classical Hamiltonian for a point 
particle in the Eddington--Robertson metric
\begin{equation}
    H^\text{ER--PPN}_\text{class} 
    = \frac{\vec p^2}{2m} + m\Phi 
        + \frac{1}{c^2} \left(-\frac{\vec p^4}{8m^3} 
            + \frac{2\gamma + 1}{2m} \vec p^2 \Phi 
            + (2\beta - 1) \frac{m \Phi^2}{2} \right) 
            + \order{c^{-4}},
\end{equation}
\end{subequations}
we see that by canonical quantisation we would not in general be able to reproduce the quantum Hamiltonian by using a specific ordering scheme, but would have to choose the ordering scheme depending on the value of $\gamma$. However, this ambiguity is due to the term in \eqref{eq:Hamiltonian_ER_WKB} proportional to $\laplacian\Phi$, which by the Newtonian gravitational field equation is proportional to the mass density of the matter generating the gravitational field. Therefore, when the quantum system is localised outside of the generating matter, we may describe it by simple canonical quantisation of the classical point-particle theory, quantising the $\vec p^2 \Phi$ term in the `obvious' symmetric ordering $\vec p \cdot \Phi \vec p$.

\begin{important}{Conclusion 2.1}%
    Formal $c^{-1}$-expansions of classical relativistic wave equations in post-Newtonian spacetimes provide a systematic method for the description of single free quantum particles in post-Newtonian gravitational background fields, motivated from the framework of QFTCS.
\end{important}

\begin{important}{Conclusion 2.2}%
    For the description of single free spin-0 quantum particles in a background post-Newtonian gravitational field as described by the Eddington--Robertson PPN metric, naive canonical quantisation of free point particle dynamics gives the same result as QFTCS-inspired post-Newtonian expansion of the Klein--Gordon equation, up to terms that vanish outside the matter distribution generating the gravitational field.
\end{important}

\subsubsection{Composite systems}
\label{subsubsec:composite-systems}

In this section, we will describe a systematic method for the derivation of a complete Hamiltonian description of an electromagnetically bound two-particle quantum system in a post-Newtonian gravitational background field described by the Eddington--Robertson PPN metric to order $c^{-2}$. 
Extending the systematic calculation by Sonnleitner and Barnett for the non-gravitational case \cite{sonnleitnerMassEnergyAnomalousFriction2018} to the gravitational situation, this provides a properly relativistic derivation of the coupling of composite quantum systems to post-Newtonian gravity, without the need to `guess' this coupling based on heuristic `relativistic effects'. We aim to keep the present discussion rather short; full details may be found in chapter 4 of \cite{schwartzPostNewtonianDescription2020} and, in a somewhat less general form, in \cite{schwartzPostNewtonianHamiltonianAtom2019}.

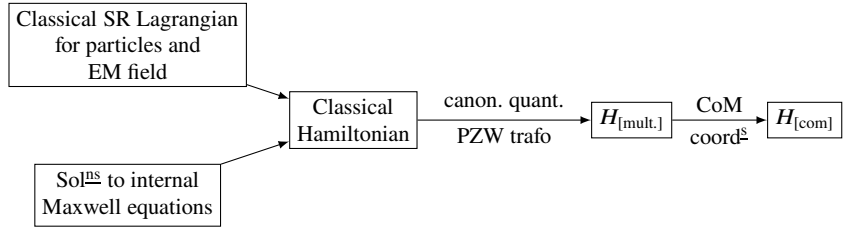
\begin{figure}[b]
    \begin{tikzpicture}
        \node[draw,align=center](classlag) at (0,1) {Classical SR Lagrangian \\ for particles and \\ EM field};
        \node[draw,align=center](maxwell) at (0,-1) {Sol$^\text{\underline{ns}}$ to internal \\ Maxwell equations};

        \node[draw,align=center](classham) at (3,0) {Classical \\ Hamiltonian};
        \draw [-latex] (classlag) -- (classham);
        \draw [-latex] (maxwell) -- (classham);

        \node[draw,align=center](hmult) at (6.7,0) {$H_\text{[mult.]}$};
        \draw [-latex] (classham) --
            node [above] {canon.\ quant.}
            node [below] {PZW trafo}
            (hmult);

        \node[draw,align=center](hcom) at (9,0) {$H_\text{[com]}$};
        \draw [-latex] (hmult) --
            node [above] {CoM}
            node [below] {coord$^\text{\underline{s}}$}
            (hcom);
    \end{tikzpicture}

    \caption{Strategy of the derivation of an `approximately relativistic' Hamiltonian for a toy hydrogenic system without gravity by Sonnleitner and Barnett in \cite{sonnleitnerMassEnergyAnomalousFriction2018}}
    \label{fig:SonnleitnerBarnetSchematic}
\end{figure}

The idea of the non-gravitational derivation by Sonnleitner and Barnett, taking place in Minkowski spacetime, is as follows \cite{sonnleitnerMassEnergyAnomalousFriction2018} (compare figure \ref{fig:SonnleitnerBarnetSchematic}). Sonnleitner and Barnett start from the classical special-relativistic Lagrangian describing two point particles of arbitrary masses and opposite and equal electric charges, interacting with electromagnetic potentials. They then split the electromagnetic potentials into `internal' (i.e.\ generated by the particles) and `external' parts, employ the Coulomb gauge, and solve the Maxwell equations for the internal part to lowest order in $c^{-1}$, expressing the solutions in terms of the particles' positions and velocities. Re-inserting the internal solutions into the classical Lagrangian\footnote
    {In the method as presented by Sonnleitner and Barnett in \cite{sonnleitnerMassEnergyAnomalousFriction2018}, a small inconsistency arises at this point. If we want to keep the external vector potential as a dynamical variable in the action, then on the one hand, its equations of motion have to be the vacuum Maxwell equations, while on the other hand, it has to enter the equations of motion of the particles. For the Lagrangian which arises from directly inserting the internal potentials, this is indeed the case: variation of the corresponding action leads to the desired equations of motion. However, this Lagrangian contains second-order time derivatives of the particle positions, such that one cannot employ conventional Hamiltonian formalism.

    Sonnleitner and Barnett disregard the second-order time derivative terms, arguing that they are related to formally diverging backreaction terms. However, \emph{this} is again problematic: these terms would have been the ones ensuring the vacuum Maxwell equations as equations of motion for the external potential; without them, the Lagrangian gives, again, the \emph{sourced} Maxwell equations for the external potential, and the formalism becomes inconsistent.

    We may avoid this inconsistency by simply treating the external electromagnetic potentials as given \emph{background} fields instead of dynamical variables. This ensures the consistency of the equations of motion while still allowing us to perform the internal--external field split. An extensive discussion of this point may be found in chapter 4 of \cite{schwartzPostNewtonianDescription2020}. Note that this inconsistency was addressed neither by Sonnleitner and Barnett in \cite{sonnleitnerMassEnergyAnomalousFriction2018} nor in our (two of the present authors') article \cite{schwartzPostNewtonianHamiltonianAtom2019}.},
expanding it to order $c^{-2}$, and performing a Legendre transformation, they arrive at a classical Hamiltonian describing the system in the post-Newtonian regime. By canonically quantising this classical Hamiltonian and performing a Power--Zienau--Woolley (PZW) unitary transformation together with a multipolar expansion of the external electromagnetic field, they then obtain a quantum `multipolar Hamiltonian' $H_\text{[mult.]}$. This describes the two particles as quantum particles, interacting with each other through the (eliminated) internal and with the external electromagnetic fields. The PZW transformation is a standard method in quantum optics, used to transform a Hamiltonian describing interactions of particles with the electromagnetic field from a minimally coupled form in terms of the potentials to a multipolar form in terms of the field strengths. Finally, Sonnleitner and Barnett introduce Newtonian centre of mass and relative coordinates and the corresponding canonical momenta, arriving at a `centre of mass Hamiltonian' $H_\text{[com]}$. This has an interpretation in terms of central and internal dynamics of a `composite particle', coupled to each other via the formal replacement $m \to M + H_\text{int}/c^2$ in the central Hamiltonian as in the heuristic motivation \eqref{eq:coupling_int_ext_heuristic}, as well as to the external electromagnetic field.\footnote
    {In addition to the `internal' part of the Hamiltonian describing the internal motion of the atom, the `central' part describing the motion of the centre-of-mass degrees of freedom together with the expected internal--external coupling, and the part describing the interaction with the external electromagnetic field, there arises a `cross term' Hamiltonian $H_\text{X}$. This contains additional couplings between the internal degrees of freedom and the central momentum. Sonnleitner and Barnett continue with the construction of a canonical transformation that eliminates these cross terms. Of course, for a thorough description of physical situations, e.g.\ in experiments, one cannot simply \emph{assume} that the resulting new coordinates correspond to the physically realised observables, but has to be careful to express all results in terms of operationally clearly defined quantities. A similar issue will become relevant in the gravitational case, as discussed below. \label{foot:cross_terms}}

We are now going to explain how to extend this method in order to include a weak gravitational background field, described by the Eddington--Robertson PPN metric~\eqref{eq:ER_PPN_metric}. Our geometric post-Newtonian expansion scheme described in the beginning of section \ref{subsec:description-model-systems}, based on the background Minkowski metric $\eta$ and the time evolution field $u$, allows us to include gravity very easily, at least conceptually speaking: in our adapted coordinates $(ct, x^a)$ we may repeat all the steps of the derivation by Sonnleitner and Barnett, the only difference being that we now start from the general-relativistic action describing our particles and fields \emph{in the curved spacetime geometry} as given by $g^\text{ER--PPN}$.
This leads to `gravitational corrections' being included in virtually all steps of the derivation, i.e.\ additional terms including the gravitational potential $\Phi$. Finally, geometric operations appearing in the resulting Hamiltonian may be re-written in terms of the physical spacetime metric, whereby some terms obtain a somewhat more intuitive interpretation in terms of metric quantities.

The action we start from, in which our particles and fields are minimally coupled to the spacetime metric $g = g^\text{ER--PPN}$, reads as follows:
\begin{align}
    S_\text{total} &= - c^2 \sum_{i=1}^2 m_i \int \D t \sqrt{-g_{\mu\nu} \dot r_i^\mu \dot r_i^\nu/c^2} 
    \nnl &\bleq
        + \int\D t \, \D^3 x \, \sqrt{-g} \left(-\frac{\varepsilon_0 c^2}{4} F_{\mu\nu} F^{\mu\nu} 
           + J^\mu A_\mu\right) \,.
\end{align}
Here, $m_i$ are the masses of the particles, $r_i^\mu(t)$ the coordinates of their worldlines, $\sqrt{-g}$ denotes the square root of minus the determinant of the matrix of metric components, $A_\mu$ are the components of the total electromagnetic four-potential, $F_{\mu\nu} = \partial_\mu A_\nu - \partial_\nu A_\mu$ are the components of the field strength, and $J^\mu = j^\mu/\sqrt{-g}$ are the components of the current four-vector field of the particles. The current density $j$, which is a vector field density, is given by
\begin{equation}
    j^\mu(t,\vec x) = \sum_{i=1}^2 e_i 
        \delta^{(3)}(\vec x - \vec r_i(t)) \dot r_i^\mu(t),
\end{equation}
where $e_1 = - e_2 =: e$ are the electric charges of the particles and $\delta^{(3)}$ is the three-dimensional Dirac delta distribution.

By inserting the Eddington--Robertson metric into the kinetic terms of the particles and expanding in $c^{-1}$, we obtain `gravitational corrections' to the kinetic terms. These may then very easily be directly included into the derivation by Sonnleitner and Barnett. For the electromagnetic fields as well, the computation does not pose any intrinsic difficulties: the Maxwell equations in the gravitational field may be written in terms of the gravity-free Maxwell equations, which allows to perturbatively include the gravitational effects on top of the `internal' potential solutions from the non-gravitational case. The derivation of this result however turns out to be quite lengthy, in particular if we do not neglect derivatives of the gravitational potential $\Phi$. Full details of the calculations may be found in chapter 4 of \cite{schwartzPostNewtonianDescription2020}.\footnote
    {Note that for simplicity, in our article \cite{schwartzPostNewtonianHamiltonianAtom2019} we neglected derivatives of $\Phi$ in the treatment of the electromagnetic fields.}

The resulting Hamiltonian takes the symbolic form\footnote
    {To simplify the presentation, here we leave out the `cross terms' $H_\text{X}$ that arise in our gravitational case exactly as in the non-gravitational case; see footnote \ref{foot:cross_terms} on page \pageref{foot:cross_terms}. We also leave out an additional cross term $\frac{2\gamma+1}{2Mc^2}[\vec P \cdot (\vec r \cdot \vec\nabla\Phi(\vec R)) \vec p_{\vec r} + \text{H.c.}]$ involving the derivative of the gravitational potential, since it is irrelevant for the following general discussion.}
\begin{subequations}
\begin{equation}
    H_\text{[com],total} = H_\text{C} + H_\text{A} 
        + H_\text{AL} + \order{c^{-4}},
\end{equation}
where $H_\text{AL}$ describes the atom-light interaction, i.e.\ the interaction of our system with the external electromagnetic fields, $H_\text{C}$ can be interpreted as describing the dynamics of the central degrees of freedom, and $H_\text{A}$ the internal atomic motion. The latter two are given by \cite{schwartzPostNewtonianDescription2020}
\begin{align}
    H_\text{C} 
    &= \frac{\vec P^2}{2M} 
            \left[1 - \frac{1}{M c^2} \left(\frac{\vec p^2}{2\mu} 
                - \frac{e^2}{4\pi\varepsilon_0 r}\right)\right] 
        + \left[M + \frac{1}{c^2} \left(\frac{\vec p^2}{2\mu} 
            - \frac{e^2}{4\pi \varepsilon_0 r}\right) \right] \Phi(\vec R) \nnl
        &\quad - \frac{\vec P^4}{8M^3 c^2} 
        + \frac{2\gamma+1}{2Mc^2} \vec P \cdot \Phi(\vec R) \vec P 
        + (2\beta-1) \frac{M \Phi(\vec R)^2}{2 c^2}, 
        \label{eq:Hamiltonian_com_C} \displaybreak[0]\\
    H_\text{A} 
    &= \left(1 + 2\gamma\frac{\Phi(\vec R)}{c^2}\right) 
            \frac{\vec p^2}{2\mu} 
        - \left(1 + \gamma\frac{\Phi(\vec R)}{c^2}\right) 
            \frac{e^2}{4\pi\varepsilon_0 r} \nnl
        &\quad - \frac{M - 3 \mu}{M} \, \frac{\vec p^4}{8 \mu^3 c^2} 
        - \frac{e^2}{4\pi\varepsilon_0} \, \frac{1}{2\mu M c^2} 
            \left( \vec p \cdot \frac{1}{r} \vec p 
            + \vec p \cdot \vec r \frac{1}{r^3} \vec r 
                \cdot \vec p \right) \nnl
        &\quad - \frac{2\gamma+1}{2 M c^2} \, \frac{\Delta m}{\mu} 
            \vec p \cdot (\vec r \cdot \vec\nabla\Phi(\vec R)) \vec p 
        + (\gamma+1) \frac{e^2}{4\pi \varepsilon_0 r} \, \frac{\Delta m}{2 M c^2}\, 
            \vec r \cdot \vec\nabla \Phi(\vec R), 
        \label{eq:Hamiltonian_com_A}
\end{align}
\end{subequations}
where $\vec R,\vec P$ are the central position and momentum, $\vec r, \vec p$ are the relative position and momentum, $M, \mu$ are the total and reduced mass of the system, respectively, and $\Delta m = m_1-m_2$ the mass difference. 
In these expressions, `dot products' of three-vectors are taken with respect to the \emph{flat} metric $\delta$ induced on three-space by the background Minkowski metric $\eta$, e.g.\ $r = \sqrt{\delta_{ab} r^a r^b}$ and $\vec p^2 = \delta^{ab} p_a p_b$.

The `physical spatial metric' $\gspac$, i.e.\ the metric on three-space induced by the \emph{physical} spacetime metric $g = g^\text{ER--PPN}$, is given as
\begin{subequations}
\begin{align}
    \gspac &= \left(1 - 2 \gamma \frac{\Phi}{c^2}\right) \delta + \order{c^{-4}}
\intertext{(compare the Eddington--Robertson metric \eqref{eq:ER_PPN_metric}), from which we see that its inverse is}
    \gspacInv &= \left(1 + 2 \gamma \frac{\Phi}{c^2}\right) \delta^{-1} + \order{c^{-4}}.
\end{align}
\end{subequations}
Thus we may rewrite the kinetic and Coulomb interaction terms from $H_\text{A}$ in \eqref{eq:Hamiltonian_com_A} in terms of $\gspac$ as
\begin{subequations} \label{eq:rewrite_int_energy_metric}
\begin{align}
    \left(1 + 2\gamma\frac{\Phi(\vec R)}{c^2}\right) 
        \frac{\vec p^2}{2\mu} 
    &= \frac{\gspacInv[\vec R] (\vec p, \vec p)}{2\mu} 
        + \order{c^{-4}} , \\
    - \left(1 + \gamma\frac{\Phi(\vec R)}{c^2}\right) 
        \frac{e^2}{4\pi\varepsilon_0 r} 
    &= - \frac{e^2}{4\pi\varepsilon_0 \sqrt{\gspac[\vec R] (\vec r, \vec r)}} 
        + \order{c^{-4}}.
\end{align}
\end{subequations}
In this form, these terms look like the usual kinetic and Coulomb interaction energies from atomic physics in Galilei-invariant QM, only with the Euclidean metric used to measure distance and momentum-squared replaced by the physical spatial metric. Thus the internal atomic Hamiltonian takes the form
\begin{subequations}
\begin{align}
    H_\text{A} &= \frac{\gspacInv[\vec R] (\vec p, \vec p)}{2\mu} 
        - \frac{e^2}{4\pi\varepsilon_0 \sqrt{\gspac[\vec R] (\vec r, \vec r)}} \nnl
        &\quad + (\text{$c^{-2}$ SR \& `Darwin' corrections} + \text{$c^{-2} \vec\nabla\Phi$ term}) 
        + \order{c^{-4}}.
\intertext{The internal--external coupling terms that we have included into the central Hamiltonian $H_\text{C}$ in \eqref{eq:Hamiltonian_com_C} may then be expressed in terms of $H_\text{A}$, leading to}
    H_\text{C} &= \frac{\vec P^2}{2M} 
            \left(1 - \frac{H_\text{A}}{M c^2} \right) 
        + \left(M + \frac{H_\text{A}}{c^2} \right) \Phi(\vec R) \nnl
        &\quad - \frac{\vec P^4}{8M^3 c^2} 
        + \frac{2\gamma+1}{2Mc^2} \vec P \cdot \Phi(\vec R) \vec P 
        + (2\beta-1) \frac{M \Phi(\vec R)^2}{2 c^2} + \order{c^{-4}}.
\end{align}
\end{subequations}
Comparing this to the Hamiltonian for a single particle in the Eddington--Robertson metric \eqref{eq:Hamiltonian_ER}, we see that this has the form
\begin{equation}
    H_\text{C} 
    = H^\text{ER--PPN}_\text{single particle}(\vec P, \vec Q, 
        M + H_\text{A}/c^2) + \order{c^{-4}}.
\end{equation}
Thus, starting from first principles our calculation supports 
the heuristic picture of a `composite point particle' whose mass is given, according to 
the `mass defect', by the total rest mass plus the 
internal energy divided by $c^2$.

\begin{important}{Conclusion 2.3}%
    Composite quantum systems in post-Newtonian gravity can be described in a systematic and properly relativistic way, starting from well-understood first principles. The results confirm, to some extent, the heuristically motivated internal--external couplings from the `mass defect' picture as in \eqref{eq:coupling_int_ext_heuristic}.
\end{important}

Note that this interpretation in terms of a `composite point particle' whose internal energy contributes to its mass---the `inertial' one in the kinetic term as well as the `gravitational' one coupling to $\Phi$---depends crucially on the rewriting \eqref{eq:rewrite_int_energy_metric} of the internal atomic energies in terms of the momentum-squared and distance as measured with the `physical spatial metric' $\gspac$. Instead, we could have chosen to interpret as internal kinetic and interaction energies the terms
\begin{subequations} \label{eq:Hamiltonian_A_parts_backgr}
\begin{align}
    H_\text{A,kin.}^\text{backgr.} 
        &= \frac{\delta^{-1}(\vec p,\vec p)}{2\mu} = \frac{\vec p^2}{2\mu} , \\
    H_\text{A,interact.}^\text{backgr.} 
        &= - \frac{e^2}{4\pi\varepsilon_0 \sqrt{\delta(\vec r, \vec r)}} 
        = - \frac{e^2}{4\pi\varepsilon_0 r} 
\end{align}
\end{subequations}
expressed with the \emph{background} spatial metric $\delta$. This would have meant to take as internal energy the part
\begin{align} \label{eq:Hamiltonian_com_A_backgr}
    H_\text{A}^\text{backgr.} &= \frac{\vec p^2}{2\mu} 
        - \frac{e^2}{4\pi\varepsilon_0 r} 
        + (\text{$c^{-2}$ SR \& `Darwin' corrections} \nnl
        &\qquad+ \text{$c^{-2} \vec\nabla\Phi$ term}) 
        + \order{c^{-4}}
\end{align}
of the total Hamiltonian. Then we would have had to interpret the left-over $\gamma$ terms from $H_\text{A}$ as part of the central Hamiltonian, giving it the form
\begin{align}
    H_\text{C}^\text{backgr.} 
    &= H_\text{C} + 2\gamma\frac{\Phi(\vec R)}{c^2} \frac{\vec p^2}{2\mu} 
        - \gamma\frac{\Phi(\vec R)}{c^2} \frac{e^2}{4\pi\varepsilon_0 r} \nnl
    &= \frac{\vec P^2}{2M} 
            \left(1 - \frac{H_\text{A}^\text{backgr.}}{M c^2} \right) 
        + \left[M + \frac{H_\text{A}^\text{backgr.}}{c^2} 
            + \frac{\gamma}{c^2} \left(2 \frac{\vec p^2}{2\mu} 
                - \frac{e^2}{4\pi\varepsilon_0 r}\right) \right] 
            \Phi(\vec R) \nnl
        &\quad - \frac{\vec P^4}{8M^3 c^2} 
        + \frac{2\gamma+1}{2Mc^2} \vec P \cdot \Phi(\vec R) \vec P 
        + (2\beta-1) \frac{M \Phi(\vec R)^2}{2 c^2} + \order{c^{-4}}.
\end{align}
Naively looking at this equation, we could have 
concluded that not only does the internal energy \eqref{eq:Hamiltonian_com_A_backgr} of the composite 
system contribute differently to the inertial and 
the (passive) gravitational masses of the composite 
particle, but also that internal kinetic and 
interaction energies \eqref{eq:Hamiltonian_A_parts_backgr} contribute differently to the (passive) gravitational 
mass. That is, we might have concluded an `anomalous coupling' of internal energies to the gravitational potential $\Phi$. Note that, due to the virial 
theorem, the time average of the additional coupling 
term $2 H_\text{A,kin.}^\text{backgr.} + H_\text{A,interact.}^\text{backgr.}$ vanishes, which for stationary states solves the apparent interpretational tension regarding the `composite point particle' 
picture~\cite{nordvedtGravInertMass1970,carlipKineticEnergyEquivalencePrinciple1998,zychGravMass2019}.

One might be tempted to argue that `real' physical distances and times as measured by `rods and clocks' are the ones as defined by the physical spacetime metric $g$, and that therefore the `correct' way to express internal energies is in terms of metric quantities with respect to $g$ (as we did before in \eqref{eq:rewrite_int_energy_metric}), which makes the `anomalous' couplings go away. However, this supposed argument is not sufficient to argue for or against the absence of such couplings in all physical situations: just because the \emph{Hamiltonian} looks natural when expressed in terms of $g$, we do not know anything about the \emph{state} of the system. The state might, a priori, have been prepared in a way that is sensitive not only to lengths as defined by $g$, but also to some other geometric structures.\footnote
    {For example, spatial light propagation in a static spacetime is described by the so-called \emph{optical metric}, which is determined by the spacetime metric $g$ and the staticity Killing field.}
For a proper analysis of physical situations, e.g.\ in experiments, one has (in principle) to describe the whole situation, including all preparation and measurement procedures, in terms of operationally defined quantities, and express all predicted results in terms of these operational quantities.

\begin{important}{Conclusion 2.4}%
    Although metric lengths are usually interpreted as being measured by `rods and clocks', they are not necessarily operationally realised in physical situations. Therefore, for a proper assessment if `anomalous' couplings of internal energies to the gravitational potential are physically relevant, it is not sufficient to show that they are eliminated from the Hamiltonian by rewriting the coupling in terms of metric lengths---a proper analysis needs to describe the \emph{whole} situation in terms of operationally defined quantities.
\end{important}

\subsection{Seemingly anomalous couplings of internal energies---a perspective from diffeomorphism invariance}
\label{subsec:anomalous-couplings}

Towards the end of the previous section, we have seen a specific example of the emergence of `anomalous' couplings of internal energies to the Newtonian potential in the systematic (post-)Newtonian description of composite systems in gravitational fields: when the internal energies are written in terms of metric quantities with respect to the background metric, defining the `absence' of gravitational fields, the `gravitational mass' multiplying the Newtonian potential in lowest order contains, in addition to the rest mass, not only a contribution of total internal energy divided by $c^2$, but an additional contribution proportional to $(2 \; \text{kinetic energy} + \text{potential energy})$.
When the internal energies are expressed in terms of metric quantities with respect to the \emph{physical} metric, instead, these `anomalous' couplings disappear from the Hamiltonian (since they are `absorbed' into the definitions of the metric lengths). We also noticed that the time average of the additional coupling term vanishes due to the virial theorem.

Such apparently `anomalous' coupling terms 
are not special to quantum-theoretic descriptions of 
composite systems. 
They typically appear in post-Newtonian descriptions 
of the dynamics of composite systems in GR. This was already noted early in the history of GR, 
like, e.g., in 1938 by Eddington and Clark in 
their approximate formulation of the dynamics of 
$n$ gravitationally interacting point masses
\cite{eddingtonClarkProblemNBodies1938}.
They found an expression for the total active
gravitational mass of the system (by looking 
at the asymptotic form of the metric) that was 
given by the sum of the individual masses, 
\emph{three} times the internal kinetic energy, 
and \emph{twice} the internal potential energy. 
This differs from the simple sum of all three 
energies (rest, kinetic, and potential) by 
$(2\;\text{kinetic energy}+\text{potential energy})$,
so as if the active gravitational mass 
did not mach the sum of all energies 
(divided by $c^2$). However, as they also 
immediately noted, and as was well known 
from the Newtonian theory of gravitating 
point masses \cite{eddingtonKineticEnergyStarCluster1916}, the apparent excess energy 
$(2\;\text{kinetic energy}+\text{potential energy})$
equals the second time-derivative of 
the total moment of inertia, which (being a total 
time derivative) clearly vanishes for bounded 
motions upon forming the time average.

In this section we aim to give more general arguments 
on the appearance of such seemingly `anomalous' couplings and the
possibility to `hide' them by coordinate redefinitions. Our exposition is based on a discussion by Carlip~\cite{carlipKineticEnergyEquivalencePrinciple1998}, in which he speaks of employing `general covariance' to argue about the `anomalous' couplings. We will make the underlying assumptions of this argumentation more explicit, in particular emphasising the importance of background structures for the arguments---namely, the background metric $\eta$ and time evolution field $u$, as introduced in the beginning of section \ref{subsec:description-model-systems}. Apart from being essential for the very definition of the notion of `weak fields', this background structure is also important when considering the matter energy-momentum tensor, which is only defined \emph{with respect to a metric}.

In order to be more precise, we will avoid the 
somewhat diffuse terminology `general covariance' and 
speak of \emph{diffeomorphism invariance} instead. 
In fact, this is more than just linguistic pedantry: 
as we will see, it is an important conceptual 
point that we consider \emph{active} diffeomorphisms
rather than just passive changes of coordinates. 
The main difference is that the latter necessarily 
affect \emph{all} fields, whereas the former may be selectively applied to \emph{some} fields, while 
leaving invariant others. This we apply, e.g., to 
keep a meaningful distinction between dynamic 
fields, which get acted upon by the diffeomorphisms, 
and background fields, which are left invariant.
Furthermore, we make precise the claim~\cite{carlipKineticEnergyEquivalencePrinciple1998} that by 
arguments based on the vanishing of `anomalous' 
coupling terms, one may prove the special-relativistic virial theorem.

Note that for the arguments in this section, 
quantisation of matter is not needed, and we 
will argue in the language of classical 
Lagrangian field theory.

\subsubsection{The mathematical setting}

Let $\mathcal M$ be the spacetime manifold. We will
consider matter fields $\Psi$ which are sections in 
some \emph{natural} vector bundle $E$ on $\mathcal M$.
We emphasise the condition of naturality, which
implies that each diffeomorphism 
$\varphi\in\mathrm{Diff}(\mathcal M)$ has a 
naturally associated lift to the total space $E$
\cite{kolarNaturalOperators1993}.
More precisely, this means that associated to each 
$\varphi\in\mathrm{Diff}(\mathcal M)$ we 
have a unique vector bundle isomorphism 
$(\hat \varphi, \varphi)\colon E \to E$
consisting of a family of linear isomorphisms
\begin{equation}
    \hat\varphi_p \colon E_p \to E_{\varphi(p)}
\end{equation}
depending smoothly on the point $p \in \mathcal M$, 
in such a way as to satisfy 
$\widehat{\mathrm{id}_{\mathcal M}} = \mathrm{id}_E$ 
and $\widehat{\varphi_1 \circ \varphi_2} = \hat \varphi_1 \circ \hat \varphi_2$. We also assume that $\hat\varphi$ depends `smoothly' on $\varphi$ in a weak sense that will be made precise below. Examples of natural bundles are 
all vector bundles associated to the (general linear) 
frame bundle of $\mathcal M$.\footnote
    {Note that a priori this excludes spinor fields, as spinor bundles do not admit natural lifts of general diffeomorphisms, but only of paths of isometries that start at the identity, being associated to the bundle of `spin frames', i.e.\ a spin structure, which is a double cover of the bundle of \emph{orthonormal} frames. Below, most of our arguments employ `infinitesimal' diffeomorphisms, i.e.\ Lie 
    derivatives of fields with respect to vector
    fields. Here the same comment applies. Lie
    derivatives of sections in $E$ only exist 
    if vector fields on $\mathcal M$ lift to vector 
    fields on the total space of the frame bundle.
    For spinor fields, this is a priori not the case,
    unless the vector field generates an isometry 
    (i.e., is a Killing field). Lie derivatives for 
    spinor fields with respect to general vector fields 
    can be defined relative to an
    extra prescription for lifting
    the vector field to the frame bundle
    \cite{kosmannDeriveesLieSpineurs1971,godinaLieDerivativeSpinor2005}.}
The space of sections $\Gamma(E)$ then carries a natural pushforward action of the diffeomorphism group,
\begin{equation}
    \mathrm{Diff}(\mathcal M) \times \Gamma(E) \ni (\varphi,\Psi) \mapsto \varphi_*\Psi := \hat\varphi \circ \Psi \circ \varphi^{-1} \in \Gamma(E).
\end{equation}
It is this pushforward that we need to be `smooth' in $\varphi$ in a weak sense: for $\varphi_t$ the local flow of any vector field $X$, we assume $(\varphi_t)_*\Psi$ to depend smoothly on $t$, such that the Lie derivative $\mathcal L_X \Psi = \left.\frac{\D}{\D t} (\varphi_{-t})_* \Psi \right|_{t=0}$ exists and we have
\begin{equation}
    (\varphi_t)_*\Psi = \Psi - t\mathcal L_X \Psi + \order{t^2}.
\end{equation}

Let now $S\colon \mathrm{Lor}(\mathcal M) \times \Gamma(E) \times \mathcal O(\mathcal M) \to \mathbb R, (g,\Psi,V) \mapsto S[g,\Psi;V]$ be the matter action, taking as inputs a Lorentzian metric $g$ on $\mathcal M$, a matter configuration $\Psi$, and an open `region of integration' $V$ in $\mathcal M$. (We do not necessarily need the action be defined as an integral over some Lagrangian density, as long as the assumptions we make in the following make sense and are satisfied. However, action integrals in this usual sense provide the most important class of actions.) Usually we will take the integration region $V$ infinite in space but finite in time, such as to render the action finite. We define the matter energy-momentum tensor by the functional derivative\footnote
    {This is simply the Hilbert energy-momentum tensor, i.e.\ the energy-momentum tensor which would appear on the right-hand side of Einstein's equations if we added the Einstein--Hilbert action as the dynamical action for the metric $g$.}
\begin{equation}
    \frac{\delta S}{\delta g_{\mu\nu}}[g,\Psi;V] =: \frac{1}{2c} \sqrt{|\det g|} \, T^{\mu\nu}[g,\Psi],
\end{equation}
i.e., by demanding that it satisfy 
\begin{align}
    S[g + \tilde g, \Psi; V] &= S[g, \Psi; V] + \frac{1}{2c} \int_V \D^4x \sqrt{|\det g|} \, T^{\mu\nu}[g,\Psi] \tilde g_{\mu\nu} + \order{\tilde g^2} \nonumber \\
    &= S[g, \Psi; V] + \frac{1}{2c} \int_V \D\mathrm{vol}_{g} \, T^{\mu\nu}[g,\Psi] \tilde g_{\mu\nu} + \order{\tilde g^2}
\end{align}
for any sufficiently small symmetric two-tensor 
$\tilde g$, where $\D\mathrm{vol}_g$ denotes the 
Riemannian volume form of $g$. Note that the assumption 
of this functional derivative being well-defined and 
independent of the integration region $V$ amounts to a 
restriction on the possible matter actions $S$---for 
example, double integrals over $V$ are excluded. We 
also assume the energy-momentum tensor to depend \emph{locally} on the matter fields, 
meaning that $\supp(T[g,\Psi]) \subseteq \supp(\Psi)$.\footnote{Recall that 
$\supp(f)$ denotes the closure of 
the set of points at which $f$ assumes a 
non-zero value. Hence, the complement of the 
support is open and consists of those points 
for which there exist open neighbourhoods 
restricted to which $f$ is identically zero. 
Now, $\supp(T[g,\Psi]) \subseteq \supp(\Psi)$ 
is equivalent to the statement that the 
complement of $\supp(\Psi)$ is contained in 
the complement of $\supp(T[g,\psi])$, i.e.\ that 
whenever $\Psi$ vanishes identically in some 
open neighbourhood, so does $\supp(T[g,\Psi])$. 
This is, e.g., the case if $T[g,\Psi](x)$ 
depends on $\Psi(x)$ and \emph{finitely} 
many derivatives of $\Psi$ at $x$.}
For example, this holds for any action that is an integral over a Lagrangian density containing at most finitely 
many derivatives of $\Psi$.

Our fundamental assumption is that the matter action be diffeomorphism invariant, i.e.\ that for any diffeomorphism $\varphi \in \mathrm{Diff}(\mathcal M)$ we have
\begin{equation}
    S[\varphi_*g, \varphi_*\Psi; \varphi(V)] = S[g, \Psi; V]
\end{equation}
for all $g,\Psi,V$.\footnote
    {As is well-known, diffeomorphism invariance of the action implies diffeomorphism covariance of the energy-momentum tensor, i.e.
    \begin{equation*}
        T[\varphi_*g, \varphi_*\Psi] = \varphi_*(T[g,\Psi]).
    \end{equation*}
    The proof of this is as follows. Let $g$ be a Lorentzian metric on $\mathcal M$, $\tilde g$ a symmetric covariant tensor field, $\Psi$ a matter field and $\varphi$ a diffeomorphism. By the definition of the energy-momentum tensor, for $\varepsilon$ sufficiently small we have
    \begin{align*}
        &\hspace{-2em} S[\varphi_*(g + \varepsilon \tilde g), \varphi_* \Psi; \varphi(V)] - S[\varphi_* g, \varphi_* \Psi; \varphi(V)] \nnl
        &= \varepsilon \frac{1}{2c} \int_{\varphi(V)} \D\mathrm{vol}_{\varphi_* g} \, T^{\mu\nu}[\varphi_* g, \varphi_* \Psi] (\varphi_* \tilde g)_{\mu\nu} + \order{\varepsilon^2} \nnl
        &= \varepsilon \frac{1}{2c} \int_{\varphi(V)} \varphi_* \Big( \D\mathrm{vol}_g \, \big(\varphi^*(T[\varphi_* g, \varphi_* \Psi])\big)^{\mu\nu} \tilde g_{\mu\nu} \Big) + \order{\varepsilon^2} \nnl
        &= \varepsilon \frac{1}{2c} \int_V \D\mathrm{vol}_g \, \big(\varphi^*(T[\varphi_* g, \varphi_* \Psi])\big)^{\mu\nu} \tilde g_{\mu\nu} + \order{\varepsilon^2}
    \end{align*}
    where we used that the pushforward commutes with tensor products and contractions, and that the Riemannian volume form of the pushforward metric $\varphi_* g$ is the pushforward of the volume form of the original metric.
    If, now, the matter action is diffeomorphism invariant, the above expression is \emph{also} equal to
    \begin{equation*}
        S[g + \varepsilon \tilde g, \Psi; V] - S[g, \Psi; V] = \varepsilon \frac{1}{2c} \int_V \D\mathrm{vol}_{g} \, T^{\mu\nu}[g, \Psi] \tilde g_{\mu\nu} + \order{\varepsilon^2}.
    \end{equation*}
    Since $\tilde g$ was arbitrary, this implies $\varphi^*(T[\varphi_* g, \varphi_* \Psi]) = T[g, \Psi]$, i.e., the covariance law $T[\varphi_* g, \varphi_* \Psi] = \varphi_*(T[g, \Psi])$.}
Note that for a theory formulated in differential-geometric terms, this will essentially always be the case, as long as no background structures enter the definition of the action $S$. 

In order to analyse the coupling of the matter fields to weak post-Newtonian gravity, we again need background structures in order to define the notions of weak gravity and small velocities. As in section \ref{subsec:description-model-systems}, we take as background spacetime Minkowski spacetime $(\mathcal M,\eta)$, and a background time evolution vector field $u$ on it that is geodesic with respect to $\eta$ and has Minkowski square $\eta(u,u) = -c^2$. As before, for convenience we choose global Lorentzian coordinates $(x^0 = ct, x^a)$ on $(\mathcal M,\eta)$ adapted to $u$, i.e.\ coordinates such that $(\eta_{\mu\nu}) = \mathrm{diag}(-1,1,1,1)$ and $u = \partial/\partial t$, in which we will work and to which we will refer all components of tensors in the following. However, the general discussion in the following will depend only on the geometric structures $\eta$ and $u$, while being independent of the choice of coordinates.

\subsubsection{The reaction to weak gravity: emergence of seemingly anomalous couplings}

We now consider a metric which is a (small) perturbation of the background Minkowski metric, $g = \eta + h$. The matter action then takes the form
\begin{equation} \label{eq:action_linear}
    S[\eta + h, \Psi; V] = S[\eta, \Psi; V] + \frac{1}{2c} \int_V \D\mathrm{vol}_\eta \, T^{\mu\nu}[\eta, \Psi] h_{\mu\nu} + \order{h^2}
\end{equation}
of a `non-gravitational' background term plus an interaction term. This is the interaction term considered in~\cite{carlipKineticEnergyEquivalencePrinciple1998}. As integration region, we consider some `sandwich' $V = I \times \mathbb R^3$ of spacetime, for a finite temporal interval $I$.

Let us now consider a perturbation $h_{\mu\nu} = -2\frac{\Phi}{c^2} \delta_{\mu\nu}$, which is 
just of the form determined by the linearised 
Einstein equations to leading order in a 
small-velocity approximation for the sourcing 
matter. The components $\delta_{\mu\nu}$
are those of the tensor field 
$^{(4)}\!\delta := \eta+2u^\flat \otimes u^\flat/c^2$, 
where $u^\flat=\eta(u,\cdot)$ is the one-form 
associated (via $\eta$) to the vector field $u$ 
that defines the reference with respect to which 
the motion of the source is considered slow. 
In background-adapted coordinates, $\delta_{\mu\nu}$ 
is just the Kronecker delta. The interaction term 
in the Lagrangian function then takes the form
\begin{subequations}
\begin{equation}
    L_\text{int}(t) = -\int_{\{ct\} \times \mathbb R^3} \D^3x \, \frac{\Phi}{c^2} \delta_{\mu\nu} T^{\mu\nu}[\eta,\Psi].
\end{equation}
Assuming that the Newtonian potential $\Phi$ is approximately constant over the extent of our system, this becomes
\begin{equation}
    L_\text{int}(t) = -\Phi(t,\bar{\vec x}) \frac{1}{c^2} \int_{\{ct\} \times \mathbb R^3} \D^3x \, \delta_{\mu\nu} T^{\mu\nu}[\eta,\Psi], 
\end{equation}
\end{subequations}
where $\bar{\vec x}$ are the spatial coordinates of some point inside the system. Thus, the expression
\begin{equation} \label{eq:grav_mass_from_action}
    M_g = \frac{1}{c^2} \int_{\{ct\} \times \mathbb R^3} \D^3x \, \delta_{\mu\nu}T^{\mu\nu}[\eta,\Psi]
\end{equation}
features as the `gravitational mass' of the system in linear order, i.e.\ the quantity coupling to the Newtonian potential.

At this point, we see how the `anomalous' coupling terms arise: the energy of our system $\Psi$, with respect to the background structures $\eta$ and $u$, is given as
\begin{equation} \label{eq:energy_from_action}
    E = \int_{\{ct\} \times \mathbb R^3} \D^3x \, T^{00}[\eta,\Psi],
\end{equation}
while the `gravitational mass' \eqref{eq:grav_mass_from_action} receives the additional contribution
\begin{equation} \label{eq:anomalous_coupling_from_action}
    M_g c^2 - E = \sum_{a=1}^3 \int_{\{ct\} \times \mathbb R^3} \D^3x \, T^{aa}[\eta,\Psi]
\end{equation}
on top of that energy. For a massive point particle on a worldline $z(t) = (ct, \vec z(t))$ in a general metric, the energy-momentum tensor is given by
\begin{equation}
    T_\text{part.}^{\mu\nu}[g,\vec z](x) = \frac{mc}{\sqrt{|\det g|}(x)} \delta^{(3)}(\vec x - \vec z(t)) \frac{\dot z^\mu(t) \dot z^\nu(t)}{\sqrt{-g(\dot z(t), \dot z(t))}} \; ,
\end{equation}
so the energy \eqref{eq:energy_from_action} and the `anomalous' term \eqref{eq:anomalous_coupling_from_action} evaluate to
\begin{subequations}
\label{eq:anomalous_coupling_particle_deriv}
\begin{align}
    E_\text{part.} &= m c^2 + \frac{m}{2} \dot{\vec z}(t)^2 + \order{c^{-2}} \nnl
        &=: mc^2 + E_\text{kin.,Newt.} + \order{c^{-2}}, \\
    (M_g c^2 - E)_\text{part.} &= 2 E_\text{kin.,Newt.} + \order{c^{-2}}.
    \label{eq:anomalous_coupling_particle}
\end{align}
\end{subequations}
\begin{question}{Exercise}\vspace{-12pt}
	\begin{prob}
		\begin{quotation}
			Using the energy-momentum tensor for a point particle, derive \eqref{eq:anomalous_coupling_particle_deriv}.
		\end{quotation}
	\end{prob}
\end{question}

In contrast, for `ultra-relativistic' matter, e.g.\ the electromagnetic field, the energy-momentum tensor is traceless, $g_{\mu\nu} T_\text{ultra-rel.}^{\mu\nu}[g,\Psi] = 0$. For Minkowski spacetime, this directly implies $T_\text{ultra-rel.}^{00}[\eta,\Psi] = \sum_{a=1}^3 T_\text{ultra-rel.}^{aa}[\eta,\Psi]$, resulting in
\begin{equation} \label{eq:anomalous_coupling_ultra-rel}
    (M_g c^2 - E)_\text{ultra-rel.} = E_\text{ultra-rel.} .
\end{equation}
Hence, for a system of slowly moving charged particles,
interacting exclusively by their electromagnetic field, equations \eqref{eq:anomalous_coupling_particle} and \eqref{eq:anomalous_coupling_ultra-rel} combine to 
$(2 \; \text{kinetic energy} + \text{potential energy})$,
which is just the `anomalous' coupling term that we 
encountered for the specific calculation in section \ref{subsubsec:composite-systems}. Following \cite{carlipKineticEnergyEquivalencePrinciple1998}
it is thus shown to be of general 
origin.

\subsubsection{Anomalous couplings and diffeomorphism
invariance}
We will now show that the apparently `anomalous' gravitational coupling term \eqref{eq:anomalous_coupling_from_action} in the 
action depends on the representative of the 
diffeomorphism equivalence class of gravitational 
fields. In other words, it can be removed by actively transforming the field of metric perturbations
by some appropriate diffeomorphism. To make this 
explicit, we consider a general diffeomorphism 
$\varphi$ by which we transform the physical metric 
$g = \eta + h$ and matter field $\Psi$. The transformed metric $\varphi_*g$ is again to be considered as a perturbation of the \emph{same} background metric 
$\eta$. Hence, $\varphi_*g =: \eta + h'$ with 
\begin{equation} \label{eq:metric_pert_trafo}
    h' = \varphi_*g - \eta = \varphi_* h + \varphi_* \eta - \eta.
\end{equation}
Note that here, as already announced before, 
it is crucial that we think of the 
diffeomorphism $\varphi$ as an active 
transformation of fields, and not just a passive 
coordinate change. Only an active 
transformation allows to transform one 
field while keeping others fixed. 

Assuming diffeomorphism invariance of the 
matter action, we obtain
%
%\begin{align} \label{eq:action_linear_transformed}
%    S[\eta + h, \Psi; V] &= S[\eta + h', \varphi_* \Psi; \varphi(V)] \nnl
%    &= S[\eta, \varphi_* \Psi; \varphi(V)] + \frac{1}{2c} \int_{\varphi(V)} \D\mathrm{vol}_\eta \, T^{\mu\nu}[\eta, \varphi_* \Psi] h'_{\mu\nu} + \order{h'^2}.
%\end{align}
\begin{multline}\label{eq:action_linear_transformed}
    S[\eta + h, \Psi; V] = S[\eta + h', \varphi_* \Psi; \varphi(V)] \\
    = S[\eta, \varphi_* \Psi; \varphi(V)] + \frac{1}{2c} \int_{\varphi(V)} \D\mathrm{vol}_\eta \, T^{\mu\nu}[\eta, \varphi_* \Psi] h'_{\mu\nu} + \order{h'^2}.
\end{multline}
Comparing to \eqref{eq:action_linear}, we see that 
under the transformation the background term has 
changed, as well as the term coupling to the metric
perturbation.

Now we specialise again to the situation considered in 
the previous section, where the initial metric 
perturbation was of the form $h_{\mu\nu} = -2\frac{\Phi}{c^2} \delta_{\mu\nu}$, with potential 
$\Phi$ approximately constant over the extent of 
the system. We define a (linearised) diffeomorphism
$\varphi$ that rescales space by a factor of 
$1 - \frac{\Phi}{c^2}$ in that part of spacetime 
containing the system, i.e.\ on $\supp(\Psi)$. 
This $\varphi$ leaves invariant each `spatial leaf' 
$\{ct\} \times \mathbb R^3 \subset V$ (as a set), 
and transforms the metric perturbation into the form\footnote
    {In coordinate-free language, we have $h' = -2\frac{\Phi}{c^2} \frac{u^\flat \otimes u^\flat}{c^2}$ on $\supp(\varphi_* \Psi)$.}
\begin{equation} \label{eq:perturb_temporal}
    h'_{\mu\nu} = -2\frac{\Phi}{c^2} \delta_\mu^0 \delta_\nu^0 \quad \text{on} \; \supp(\varphi_* \Psi) .
\end{equation}
Put differently, application of $\varphi$ `transforms 
away' all components of the metric perturbation 
apart from the $00$~component over the extent of 
the system. Note that we have to assume that 
$\Phi$ be (approximately) constant over the system 
not only in space, but also in the temporal 
direction: if it changed over time, the diffeomorphism
would have to scale space differently at different 
times, thereby introducing mixed spatio-temporal 
terms $h'_{0a}$ in the metric perturbation.%
\footnote{In detail, the (linearised) diffeomorphism
$\varphi$ accomplishing this transformation can in 
our setting be described as follows (we refrain from 
giving coordinate-independent constructions):
Consider the vector field with components
\begin{equation*} 
%\label{eq:scaling_vector_field}
X^0 = 0,\quad X^a = -\frac{\Phi(\bar{\vec x})}{c^2} 
\xi(\vec x) (x^a - \bar x^a) \,.
\end{equation*}
Here $\bar x^a$ are the coordinates of a fixed
spatial reference position within the body,
i.e.\ within $\supp(\Psi)$, and $\xi$ is a 
function that is constantly $1$ on $\supp(\Psi)$  
and falls off rapidly to zero outside. 
Defining $\varphi$ to be the diffeomorphism 
generated by $X$ (i.e.\ the flow for unit time) 
and writing $\varepsilon := \Phi(\bar{\vec x})/c^2$ 
for brevity, by the definition of the Lie derivative 
we have
    \begin{align*}
        h' + \eta &= \varphi_*(\eta + h) \nnl
        &= \eta + h - \mathcal L_X(\eta + h) + \order{\varepsilon^2} \nnl
        &= \eta + h - \mathcal L_X\eta + \order{\varepsilon^2} \,,
    \end{align*}
    since $h$ itself is of order $\varepsilon$. This means $h' = h - \mathcal L_X\eta$ (up to higher order terms), or in components
    \begin{align*}
        h'_{\mu\nu} &= h_{\mu\nu} - \partial_\mu X_\nu - \partial_\nu X_\mu \, .
    \end{align*}
    By our choice of $X$ 
    %\eqref{eq:scaling_vector_field}, 
    the new metric perturbation takes on the desired 
    form \eqref{eq:perturb_temporal} (up to terms of 
    order $\varepsilon^2$) on $\supp(\Psi)$. 
    To conclude that it has this form also on 
    $\supp(\varphi_*\Psi) = \varphi(\supp(\Psi))$, 
    note that for any function $f$ we have 
    $f\circ\varphi = f + \mathcal L_X f + 
    \order{\varepsilon^2}$, such that
    \begin{equation*}
        f = \order{\varepsilon^2} \; \text{on} \; U 
        \implies f = \order{\varepsilon^2} \; \text{on} \; \varphi(U).
    \end{equation*}
    Applying this to $U = \supp(\Psi)$ and $f = h_{\mu b}$, we obtain \eqref{eq:perturb_temporal}.}

At this point, the assumption of locality of the energy-momentum tensor in its matter-field argument 
comes into play: due to this locality, the transformed perturbation has the `purely temporal' form \eqref{eq:perturb_temporal} also on $\supp(T[\eta, \varphi_* \Psi]) \subseteq \supp(\varphi_* \Psi)$. Therefore, transforming by $\varphi$, the action \eqref{eq:action_linear_transformed} may be written in the form
\begin{equation}
    S[\eta + h, \Psi; V] = S[\eta, \varphi_* \Psi; V] - \frac{1}{c} \int_V \D\mathrm{vol}_\eta \, \frac{\Phi}{c^2} T^{00}[\eta, \varphi_* \Psi] + \order{h'^2},
\end{equation}
and the transformed interaction term is
\begin{equation}
    L_\text{int,new}(t) = -\Phi(t,\bar{\vec x}) \frac{1}{c^2} \int_{\{ct\} \times \mathbb R^3} \D^3x \, T^{00}[\eta, \varphi_* \Psi].
\end{equation}
The quantity that now appears as the `gravitational 
mass' (after the transformation by $\varphi$) is the
spatial integral of the energy density $T^{00}[\eta,\varphi_* \Psi]$ of the transformed 
system on the Minkowski background. 
We stress once more that the background fields 
$\eta$ and $u$ are always the same, so that in 
particular the notion of `energy density' for the 
original and the transformed fields is the same, 
namely $T^{00}=T(u^\flat/c,u^\flat/c)$.

We see that in in the new diffeomorphism-equivalent representation an `anomalous' coupling as in \eqref{eq:grav_mass_from_action} no longer exists. 
Moreover, we note that in the new representation 
the spatial part of the metric is just of 
Euclidean form. 
Thus, the new representation expresses the 
quantities which describe the state of the system 
in a form in which the spatial metric in the 
rest-frame of the observer $u$ is Euclidean. 
This is precisely the representation which we 
observed in the concrete example of section \ref{subsubsec:composite-systems} to eliminate the disturbing coupling terms from the Hamiltonian. 
Hence we see that our previous example may be 
considered as a special case of a general law,
according to which apparently `anomalous' 
couplings disappear in particular, metrically
preferred representations.\footnote
{`Metrically preferred' means in our case 
that the pushforward $\varphi_*$ transforms 
the physical spatial metric 
$\gspac = g|_{\{t = \text{const.}\}}$ 
into the flat background spatial metric 
$\delta = \eta|_{\{t = \text{const.}\}}$, 
such that the $\eta$-length of a transformed 
spacelike vector $\varphi_* v$ is the same 
as the $g$-length of the original vector $v$.}

\subsubsection{Dynamical consequences of diffeomorphism invariance: the virial theorem}

Carlip~\cite{carlipKineticEnergyEquivalencePrinciple1998} 
claimed that his discussion of the elimination of the seemingly anomalous 
coupling terms by `general covariance' might even be seen as 
a \emph{derivation} of the special-relativistic virial theorem: 
since by the application of spatial rescaling diffeomorphisms as in the previous section we may arbitrarily alter the term coupling $\int \D^4x \sum_{a=1}^3 T^{aa}$ to the Newtonian potential, but the `physical coupling' ought not depend on our `choice of 
gauge', we are supposed to conclude that this 
integral must vanish. This is just the statement 
of the special-relativistic virial theorem, a discussion 
of which may be found in 
\cite[\S\,34]{landauLifshitzClassicalTheoryFields1980}.

This argument seems too good to be true: it should be clear that a statement such as the virial theorem cannot be derived from kinematical assumptions (such as diffeomorphism invariance of the action) alone; some \emph{dynamical} assumption is needed (usually, one takes local energy-momentum conservation, also called `closedness' of the system under consideration, i.e.\ $\partial_\mu T^{\mu\nu} = 0$). Looking at the transformation behaviour \eqref{eq:action_linear_transformed} of the matter action in weak gravity under the action of a diffeomorphism, we see that the proposed argument breaks down due 
to the fact that not only the coupling term but 
also the `background term' changes,
from $S[\eta,\Psi;V]$ in \eqref{eq:action_linear} 
to $S[\eta, \varphi_* \Psi; \varphi(V)]$ in \eqref{eq:action_linear_transformed}. 
Therefore, without any further assumptions, we 
cannot deduce equality of the two coupling terms.

However, the desired conclusion (virial theorem) 
can be drawn if 
\begin{equation} \label{eq:field_eq}
    \frac{\delta S}{\delta\Psi}[\eta,\Psi;V] = 0 \; \text{under variations vanishing on} \; \partial V,
\end{equation}
i.e.\ if we assume that the matter field configuration \emph{$\Psi$ solve the equations of motion} (on the Minkowski background). The precise argument runs as 
follows: Fix an $\varepsilon > 0$, and let $X$ be a 
vector field on $V$ that vanishes on the boundary 
$\partial V$. Let $\varphi$ be the flow-diffeomorphism associated to $X$ for flow parameter $\varepsilon$. 
Writing $\varphi_*\Psi = \Psi - \varepsilon \mathcal L_X \Psi + \order{\varepsilon^2}$, we see that to leading order in $\varepsilon$ the difference $\varphi_*\Psi - \Psi$ vanishes on $\partial V$, such that the field equation \eqref{eq:field_eq} implies
\begin{equation} \label{eq:action_background_equality}
    S[\eta,\varphi_*\Psi;V] 
    = S[\eta,\Psi;V] + \order{\varepsilon^2}.
\end{equation}
Expanding in $\varepsilon$, we also directly obtain
\begin{equation} \label{eq:energy--mom_trafo_linear}
    T^{\mu\nu}[\eta,\varphi_*\Psi] 
    = T^{\mu\nu}[\eta,\Psi] + \order{\varepsilon}.
\end{equation}
On the other hand, combining \eqref{eq:action_linear} and \eqref{eq:action_linear_transformed}, we have
\begin{align} \label{eq:action_trafo_equality}
    &S[\eta, \Psi; V] + \frac{1}{2c} \int_V \D\mathrm{vol}_\eta \, T^{\mu\nu}[\eta, \Psi] h_{\mu\nu} + \order{h^2} \nnl
    &= S[\eta, \varphi_* \Psi; V] + \frac{1}{2c} \int_V \D\mathrm{vol}_\eta \, T^{\mu\nu}[\eta, \varphi_* \Psi] h'_{\mu\nu} + \order{h'^2},
\end{align}
where $h'$ is given by \eqref{eq:metric_pert_trafo} in terms of $\eta$, $h$ and $\varphi$. Assuming $h$, and therefore also $h'$, to be of order $\varepsilon$, we may combine \eqref{eq:action_trafo_equality} with equality of the background terms up to quadratic terms in $\varepsilon$ \eqref{eq:action_background_equality} and equality of the energy-momentum tensor up to linear terms \eqref{eq:energy--mom_trafo_linear}, and obtain
\begin{equation} \label{eq:integral_T_pert_difference}
    0 = \int_V \D\mathrm{vol}_\eta \, T^{\mu\nu}[\eta,\Psi] (h_{\mu\nu} - h'_{\mu\nu}).
\end{equation}

We now want to apply this equation to the situation considered in the previous section, taking $V = [0,ct] \times \mathbb R^3$, $h_{\mu\nu} = -2\frac{\Phi}{c^2} \delta_{\mu\nu}$ with a constant $\Phi$, and $\varphi$ a diffeomorphism that scales space on $\supp(\Psi)$, such as to obtain $h'_{\mu\nu} =  -2\frac{\Phi}{c^2} \delta_\mu^0 \delta_\nu^0$ there. If we were able to apply \eqref{eq:integral_T_pert_difference} to these ingredients, we could conclude a statement about $\sum_{a=1}^3 T^{aa}[\eta,\Psi]$. However, we are met with an obstacle: in the derivation of \eqref{eq:integral_T_pert_difference}, we needed that the vector field generating the diffeomorphism vanish on the boundary $\partial V$. At the spatial boundary of $V$, i.e.\ at `spatial infinity', this does not pose a problem if we assume that $\Psi$ have spatially compact support---we then need $\varphi$ to scale space only in a finite spatial region, and may take its generator $X$ to fall rapidly to zero outside. At the temporal boundary of $V$ however, we apply the same spatial rescaling as at all other times, and the generating vector field does not vanish. Therefore, in our situation, \eqref{eq:integral_T_pert_difference} only holds up to a boundary term on the temporal boundary of $V$, arising from the failure of \eqref{eq:action_background_equality}. If, however, we assume this boundary term to be local in $\Psi$, it vanishes when we take the average over larger and larger time intervals (since $\supp(\Psi)$ is spatially compact). Thus, we arrive at the special-relativistic virial theorem:
\begin{equation}
    \lim_{t\to\infty} \frac{1}{t} \int_{[0,ct] \times \mathbb R^3} \D^4x \sum_{a=1}^3 T^{aa}[\eta,\Psi] = 0 \,.
\end{equation}
We thus have shown that by a suitable adaptation of Carlip's argument from \cite{carlipKineticEnergyEquivalencePrinciple1998}, one can indeed prove the virial theorem.

Note that we may, in fact, easily deduce local energy-momentum conservation from \eqref{eq:integral_T_pert_difference}: According to \eqref{eq:metric_pert_trafo}, for $h = \order{\varepsilon}$ we have in general
\begin{equation}
    h_{\mu\nu} - h'_{\mu\nu} 
    = \varepsilon (\mathcal L_X \eta)_{\mu\nu} + \order{\varepsilon^2} 
    = 2\varepsilon \partial_{(\mu} X_{\nu)} + \order{\varepsilon^2} \,.
\end{equation}
Thus, for vector fields $X$ vanishing on the boundary $\partial V$, \eqref{eq:integral_T_pert_difference} implies
\begin{align}
    0 &= \int_V \D\mathrm{vol}_\eta \, T^{\mu\nu}[\eta,\Psi] \partial_{(\mu} X_{\nu)} \nnl
    &= \int_V \D\mathrm{vol}_\eta \, T^{\mu\nu}[\eta,\Psi] \partial_\mu X_\nu \nnl
    &= - \int_V \D\mathrm{vol}_\eta \, \partial_\mu T^{\mu\nu}[\eta,\Psi] X_\nu \; ,
\end{align}
which, due to $X$ being arbitrary, implies $\partial_\mu T^{\mu\nu}[\eta,\Psi] = 0$.

\begin{important}{Conclusion 2.5}%
    Seemingly anomalous coupling terms of internal energies of a composite system to the Newtonian gravitational potential depend on the chosen representative of the 
    diffeomorphism equivalence class of fields.
    In other words, they are gauge dependent and 
    may be eliminated from the action by choosing 
    an appropriate representative (i.e.\ by `choosing a gauge'). 
    However, as already stressed in the discussion 
    in section \ref{subsubsec:composite-systems},
    %particularly in Conclusion 2.4, 
    this is not sufficient to argue for or against the \emph{physical} relevance of such coupling terms: 
    in a sense, the coupling terms have simply been `hidden' by the field redefinition $\Psi \to \varphi_*\Psi$. The question of \emph{operational} significance in concrete physical situations is still the important one, which cannot be answered by simple `kinematic' arguments as the one discussed in this section, without any assumptions about the complete physical situation.\footnote{In that respect we contradict 
    the immediate conclusion drawn in \cite[p.\,6]{zychGravMass2019}, 
    that `correctly defining internal energies 
    yields the true and unique gravitational mass and exposes the validity of the equivalence principle'.
    We also contradict the implicit statement made in 
    this quotation, namely that the validity of the 
    equivalence principle hinges on the equality of 
    various notions of `mass', the definitions of which 
    are---as we have just seen---gauge dependent.}

    Notwithstanding the non-viability of such arguments for answering questions about the physical relevance of such couplings, by making the additional \emph{dynamical} assumption of the field solving the equations of motion, arguments based on diffeomorphism invariance may be used to prove the special-relativistic virial theorem.
\end{important}

%%%%%%%%%%%%%%%%%%%%%%%%%%%%%%%%%%%%%%%%%%%%%%%%%%%%%%%%%%%%%%%%%%%%%%
%%%%%%%%%%%%%%%%%%%%%%%%%%%%%%%%%%%%%%%%%%%%%%%%%%%%%%%%%%%%%%%%%%%%%%

\section{Quantum matter with gravitational backreaction}
\label{sec:qm-backreaction}
The objective of most research in `quantum gravity' focuses on the question \emph{how} gravity can be quantised (and the consequences of this endeavour,
as discussed in this volume's Chapters on string-inspired effective modified gravity theories and on deformed relativistic symmetry principles). Taking a step back, before asking for the \emph{how}, one may first ask \emph{if} the gravitational field should be quantised at all.
Of course, this presupposes a reasonable definition of what it means for gravity to be quantised---which is the defining feature that makes a theory a quantum theory?

Einstein's equations~\eqref{eqn:einstein-eq} \emph{can} be understood as describing a \emph{field} $g_{\mu\nu}(x)$ on spacetime. From this point of view, one may find it reasonable to apply the same quantisation rules to solutions $g_{\mu\nu}(x)$ of Einstein's equations that one applies, for instance, to the classical solutions $\psi$ of the Dirac equation in order to arrive at fermionic quantum fields. However, $g_{\mu\nu}(x)$ is not simply a field \emph{on} spacetime; it describes the metric and further differential-geometric properties
of spacetime itself. Einstein's equations can only be understood as equations of a field \emph{on} spacetime in a perturbative sense: by separating the metric $g_{\mu\nu} \to g_{\mu\nu} + h_{\mu\nu}$ into some background metric $g_{\mu\nu}$ and only treating the variation $h_{\mu\nu}$ with respect to that background as a field living on the \emph{background} defined by $g_{\mu\nu}$. 

From a more philosophical point of view, one may ask whether such an artificial splitting of the structure of spacetime into background and field is a more plausible approach (even if in the end physical predictions would turn out to be independent of the way in which the splitting is done) than the alternative that gravity is fundamentally different and spacetime cannot simply be quantised in the same way as matter fields. In any case, the perturbative quantisation in the \emph{exact} same way as for matter fields cannot be ultimately correct, because it is known to result in non-renormalisable divergences~\cite{thooftOneLoopDivergencies1974,goroffQuantumGravityTwo1985}.

We can avoid the ambiguity about what it means to quantise gravity altogether by asking the opposite: can we construct a theory that consistently combines (classical) GR with quantum matter, specifically, that solves the problem of defining the right-hand side in Einstein's equations from quantum matter fields? Any theory that accomplishes this, we want to refer to as \emph{semiclassical gravity}.

\subsection{Semiclassical gravity sourced by mean energy-momentum}
\label{subsec:semiclassical-gravity}
Once a proper mathematical model for quantum fields in curved spacetime has been established, one can define an energy-momentum operator $\hat{T}_{\mu\nu}$ via canonical quantisation of the corresponding classical object. An obvious way to include the gravitational backreaction of these fields are the semiclassical Einstein equations~\cite{mollerTheoriesRelativistesGravitation1962,rosenfeldQuantizationFields1963}
\begin{equation}\label{eqn:sce}
    R_{\mu\nu} - \frac{1}{2} R \, g_{\mu\nu} = \frac{8 \pi G}{c^4} \ev{\hat{T}_{\mu\nu}} \,,
\end{equation}
where $\ev*{\hat{T}_{\mu\nu}} = \bra{\Psi} \hat{T}_{\mu\nu} \ket{\Psi}$ denotes the expectation value in the state $\ket{\Psi}\in \mathcal{F}_\pm(\mathcal{H})$ of the matter field. This choice of right-hand side ensures the correct classical limit to Einstein's equations~\eqref{eqn:einstein-eq}.

The proper Hilbert space structure for modelling quantum states in curved spacetime provided, the expectation value is readily defined. The classical energy-momentum tensor, however, is generally quadratic in the fields. A straightforward substitution of classical fields by field operators $\phi \to \hat{\phi}$ results in an object $\hat{T}_{\mu\nu}(x)$ which includes two-point correlation functions such as $\ev*{\hat{\phi}(x)\hat{\phi}(x')}$ at the \emph{same} spacetime point $x = x'$, which diverge. Therefore, in addition to the classical definition of $T_{\mu\nu}$ an appropriate renormalisation procedure is required, which results in a certain ambiguity of $\hat{T}_{\mu\nu}(x)$. It has been shown by Wald~\cite{waldBackReactionEffect1977,waldTraceAnomalyConformally1978,waldQuantumFieldTheory1994} that the renormalised energy-momentum operator compatible with the semiclassical Einstein equations~\eqref{eqn:sce} can be defined in an axiomatic way, requiring:
\begin{enumerate}
    \item For any two \emph{orthogonal} states $\braket{\Psi}{\Phi} = 0$ the matrix elements $\bra{\Psi} \hat{T}_{\mu\nu} \ket{\Phi}$ agree with those obtained by the formal substitution $\phi \to \hat{\phi}$ of classical fields with field operators in the classical energy-momentum tensor.
    \item In flat Minkowski spacetime the renormalised $\hat{T}_{\mu\nu}$ reduces to the normal ordered energy-momentum operator obtained by substituting $\phi \to \hat{\phi}$.
    \item Expectation values of the renormalised energy-momentum operator are covariantly conserved: $\nabla^\mu \ev*{\hat{T}_{\mu\nu}} = 0$.
    \item Causality holds in the sense that only changes in the metric in the causal past of some spacetime point $p$ can affect the value of $\ev*{\hat{T}_{\mu\nu}}$ at $p$.
\end{enumerate}
The renormalised $\hat{T}_{\mu\nu}$ satisfying these axioms is uniquely determined up to the addition of local curvature terms.

\subsubsection{The nonrelativistic limit of semiclassical gravity}
In order to arrive at a nonrelativistic\footnote
{Note that here we employ the common, yet somewhat misleading, adjective `nonrelativistic' to designate Galilei-invariant dynamical laws in distinction from `relativistic' ones, which then are those obeying Poincar\'e invariance. Nevertheless, we want to emphasise that it is not the validity of the physical relativity principle that distinguishes both cases; rather, their difference lies in the way in which that principle is implemented.}
\schr\ equation for semiclassical gravity, one can follow the usual procedure to derive the Newtonian limit of Einstein's equations via the linearised theory, combined with the assumption of slow velocities of the sourcing matter~\cite{bahramiSchrodingerNewtonEquationIts2014}. The metric is written as $g_{\mu\nu} = \eta_{\mu\nu} + h_{\mu\nu}$ with a perturbation $h_{\mu\nu}$ around flat Minkowski spacetime. Introducing the trace-reversed metric $\overline{h}_{\mu\nu} = h_{\mu\nu} - \tfrac12 \eta_{\mu\nu} \eta^{\rho\sigma} h_{\rho\sigma}$ and applying the de Donder gauge condition $\partial^\nu \overline{h}_{\mu\nu} = 0$, Einstein's equations to linear order in the metric perturbation yield the wave equations
\begin{equation}
    \Box \overline{h}_{\mu\nu} = -\frac{16 \pi \, G}{c^4} \ev{\hat{T}_{\mu\nu}} \,,
\end{equation}
$\Box = \nabla^\mu \nabla_\mu$ being the d'Alembert operator. In the weak field, nonrelativistic limit, the behaviour is dominated by the $00$-component, and the d'Alembert operator can be approximated by the flat space Laplace operator $\laplacian$, neglecting time derivative terms of order $c^{-2}$. Defining the Newtonian potential $\Phi = -\tfrac{c^2}{4} \overline{h}_{00}$ and the mass density operator $\hat{\rho} = \hat{T}_{00}/c^2$, one finds the Poisson equation
\begin{equation}\label{eqn:sc-poisson}
    \laplacian \Phi = 4 \pi \, G \, \ev{\hat{\rho}} \,.
\end{equation}
For a single field of mass $m$ particles, we can define the $N$-particle state
\begin{equation}
    \ket{\Psi_N} = \frac{1}{\sqrt{N!}} \int \D^3 x_1 \cdots \D^3 x_N
    \Psi_N(t,\vec x_1,\dots,\vec x_N) \hat{\psi}^\dagger(\vec x_1) \cdots \hat{\psi}^\dagger(\vec x_N) \ket{0}
\end{equation}
with the nonrelativistic field operators $\hat{\psi}$ and $N$-particle wave function $\Psi_N$. The mass density operator\footnote{Note that, contrary to perturbatively quantised gravity, all expressions derived from $\hat{\rho}$ in semiclassical gravity are well-defined, and there is no need for renormalisation.} $\hat{\rho}(\vec x) = m \hat{\psi}^\dagger(\vec x) \hat{\psi}(\vec x)$ has the time dependent expectation value
\begin{align}
    \ev{\hat{\rho}(\vec x)}_N
    &= \bra{\Psi_N} m \hat{\psi}^\dagger(\vec x) \hat{\psi}(\vec x) \ket{\Psi_N} \nnl
    &= m \sum_{i=1}^N \int \left(\prod_{j=1}^N \D^3 x_j\right)\,
    \delta^{(3)}(\vec x - \vec x_i)\,\abs{\Psi_N(t,\vec x_1,\dots,\vec x_N)}^2\,,
\end{align}
and specifically $\ev{\hat{\rho}} = m \abs{\psi}^2$ for a single particle with wave function $\psi(t,\vec x)$. Integrating the Poisson equation~\eqref{eqn:sc-poisson} results in the potential
\begin{align}\label{eqn:sc-newton-pot}
    \Phi(t,\vec x)
    &= -G\,\int \D^3 x' \frac{\ev{\hat{\rho}(\vec x')}}{\abs{\vec x - \vec x'}} \nnl
    &= -G\,m \sum_{i=1}^N \int \left(\prod_{j=1}^N \D^3 x_j\right)\,
    \frac{\abs{\Psi_N(t,\vec x_1,\dots,\vec x_N)}^2}{\abs{\vec x - \vec x_i}} \,.
\end{align}
Given the classical spacetime structure corresponding to the Newtonian gravitation potential~\eqref{eqn:sc-newton-pot}, the problem is that of section~\ref{sec:qm-class-backgrounds}: what is the \schr\ equation that follows for the dynamics of matter in said classical spacetime?

The plausible answer, confirmed for the external homogeneous potential in the earth's gravity~\cite{colellaObservationGravitationallyInduced1975}, is that the potential should enter the Hamiltonian in the usual way, i.e.\ the Hamilton operator for the evolution of $N$ particles in the position basis should be
\begin{equation}
    \hat{H}_N = \sum_{i=1}^N \left(\frac{\hat{\vec p}_i^2}{2 m} + m \Phi(t,\vec x_i)\right) + V_\text{matter} \,,
\end{equation}
where $\hat{\vec p} = -\I\hbar\nabla$ is the momentum operator and the potential $V_\text{matter}$ contains all external and internal non-gravitational forces.
The resulting \schr\ equation $\I \hbar \partial_t \Psi_N = \hat{H}_N \Psi_N$ is called the ($N$-particle) \emph{\schr--Newton} equation. Due to the wave function dependence of the gravitational potential~\eqref{eqn:sc-newton-pot}, it is a nonlinear \schr\ equation which for the case $N=1$ of a single particle reads
\begin{equation}\label{eqn:sne-single-part}
    \I \hbar \partial_t \psi(t,\vec x) = \left[
    -\frac{\hbar^2}{2m} \laplacian
    - G m^2 \int \D^3 x' \frac{\abs{\psi(t,\vec x')}^2}{\abs{\vec x - \vec x'}}
    + V_\text{ext} \right] \psi(t,\vec x) \,.
\end{equation}
The gravitational term describes a self-interaction: the particle is attracted by a distribution of its mass $m$ with the probability density $\abs{\psi}^2$. Furthermore, the balance between this gravitational self-attraction and the free spreading of the \schr\ equation results in the existence of stationary solutions~\cite{morozSphericallysymmetricSolutionsSchrodingerNewton1998}.

Despite its nonlinearity, the \schr--Newton equation maintains many of the typical properties from linear QM. Specifically, the norm of the wave function is conserved, $\partial_t \int \D^3 x \abs{\psi(t,\vec x)}^2 = 0$, allowing for a probabilistic interpretation as in standard QM. Note also that at any given time the uncertainty relation between non-commuting observables remains intact. Arguments that semiclassical gravity would violate position-momentum uncertainty~\cite{eppleyNecessityQuantizingGravitational1977}, therefore, do not apply to semiclassical gravity based on the semiclassical Einstein equations. In any case, a violation of the uncertainty relation would only constitute a testable deviation from standard QM and not an inconsistency, as long as its magnitude is not in contradiction to experimentally established values.

\begin{question}{Exercise}\vspace{-12pt}
	\begin{prob}
		\begin{quotation}
			The single particle \schr--Newton equation can be derived from 
			the Lagrangian density
			\begin{equation*}
			\mathcal{L} =
            \frac{\I\hbar}{2} \left(\psi^* \partial_t \psi - \psi \partial_t \psi^* \right)
            - \frac{\hbar^2}{2m} \abs{\nabla \psi}^2
            - V_\text{ext} \abs{\psi}^2
            - \frac{m}{2} \Phi \abs{\psi}^2
            \end{equation*}
            by variation with respect to the independent variables $\psi$ and 
            $\psi^*$, taking into account that $\Phi$ is itself a convolution
            $\Phi = U \ast \abs{\psi}^2$ of the potential 
            $U(\vec x) = -G m/\abs{\vec x}$ with the magnitude squared of the
            wave function. Show this, and that the conserved Noether charge
            for the symmetry under phase transformations
            $\psi \to \E^{\I \alpha} \psi$ is the norm of the wave function.
		\end{quotation}
	\end{prob}
\end{question}

It is sometimes claimed~\cite{anastopoulosProblemsNewtonSchrodingerEquations2014,huSemiclassicalStochasticGravity2020} that the \schr--Newton equation would not follow as the weak field nonrelativistic limit of the semiclassical Einstein equations. The criticism is based on the observation that in analogy with quantum electrodynamics one would expect the nonrelativistic, second-quantised Hamiltonian acting on Fock space states to take the form
\begin{equation}\label{eqn:fock-ham-qg}
    \hat{H}_\text{qg} = -\frac{\hbar^2}{2m} \int \D^3 x \, \hat{\psi}^\dagger(\vec x) \laplacian \hat{\psi}(\vec x)
    - G \iint \D^3 x \, \D^3 x' \, \frac{\hat{\rho}(\vec x) \hat{\rho}(\vec x')}{\abs{\vec x - \vec x'}} \,.
\end{equation}
This Hamiltonian results in divergent matrix elements which can be cured either by the introduction of a regularised mass density operator~\cite{anastopoulosProblemsNewtonSchrodingerEquations2014} $\hat{\rho}_\text{reg}(\vec x)$, smeared over some spatial region around $\vec x$, or by replacing the product of mass density operators by its normal ordered equivalent ${:\,}\hat{\rho}(\vec x) \hat{\rho}(\vec x'){\,:} = m^2 \hat{\psi}^\dagger(\vec x) \hat{\psi}^\dagger(\vec x') \hat{\psi}(\vec x) \hat{\psi}(\vec x')$.
This procedure removes the self-interaction\footnote{Exactly as in quantum electrodynamics, the self-interaction does not appear at tree level but is reintroduced via higher loop orders.} at the level of single particles, and yields a linear \schr\ equation. This is, in fact, the Hamiltonian one would expect from quantised gravity. With respect to semiclassical gravity, however, assuming an analogy to quantum electrodynamics amounts to circular reasoning. The two occurrences of $\hat{\rho}$ in equation~\eqref{eqn:fock-ham-qg} stem from the nonrelativistic limit of the coupling $\sim h_{\mu\nu} T^{\mu\nu}$ in the linearised combined action for gravity and matter. From a quantum gravity perspective, one expects both terms to be subject to canonical quantisation. In the semiclassical theory, on the other hand, one would only quantise the matter part $T^{\mu\nu}$, whereas the metric perturbation remains classical. Instead of the Hamiltonian~\eqref{eqn:fock-ham-qg} one has~\cite{bahramiSchrodingerNewtonEquationIts2014}
\begin{equation}\label{eqn:fock-ham-sc}
    \hat{H}_\text{sc} = -\frac{\hbar^2}{2m} \int \D^3 x \, \hat{\psi}^\dagger(\vec x) \laplacian \hat{\psi}(\vec x)
    - G \iint \D^3 x \, \D^3 x' \, \frac{\hat{\rho}(\vec x) \ev{\hat{\rho}(\vec x')}}{\abs{\vec x - \vec x'}} \,,
\end{equation}
with the first-quantised potential~\eqref{eqn:sc-newton-pot}.

Although the proper derivation of a nonrelativistic \schr\ equation from quantum fields in curved spacetime is an unsolved question (cf.\ section~\ref{sec:qm-class-backgrounds}), the true conflict that makes many question the validity of the \schr--Newton equation is not its derivability from the semiclassical Einstein equations. Reading between the lines, one finds that what the critique~\cite{anastopoulosProblemsNewtonSchrodingerEquations2014,huSemiclassicalStochasticGravity2020} of the \schr--Newton equation is actually based on are its `problematic consequences' with regard to its connection to the observed reality.

\subsection{Consistency of semiclassical gravity}
\label{subsec:consistency-semiclassical-gravity}
Historically, the question whether semiclassical gravity is consistent already concerned physicists in the early days of quantum field theory. During the Chapel Hill conference~\cite{chapelhill1957}, Feynman proposed the following thought experiment:
\begin{quotation}
    Suppose we have an object with spin which goes through a Stern-Gerlach experiment. Say it has spin 1/2, so it comes to one of two counters.
    Connect the counters by means of rods, etc., to an indicator which is either up when the object arrives at counter 1, or down when the object arrives at counter 2. Suppose the indicator is a little ball, 1\,cm in diameter.
    
    Now, how do we analyze this experiment according to quantum mechanics? We have an amplitude that the ball is up, and an amplitude that the ball is down. That is, we have an amplitude (from a wave function) that the spin of an electron in the first part of the equipment is either up or down. And if we imagine that the ball can be analyzed through the interconnections up to this dimension ($\approx 1$\,cm) by the quantum mechanics, then before we make an observation we still have to give an amplitude that the ball is up and an amplitude that the ball is down. Now, since the ball is big enough to produce a \emph{real} gravitational field (we know there's a field there, since Coulomb measured it with a 1\,cm ball) we could use that gravitational field to move another ball, and amplify that, and use the connections to the second ball as the measuring equipment.
\end{quotation}
Denoting with $\ket{\up}$ and $\ket{\dn}$ the spin eigenstates and with $\ket{U}$ and $\ket{D}$ the corresponding final states of the ball, the experiment amounts to creating an entangled state
\begin{equation}\label{eqn:pg-entangled}
    \ket{\Psi} = \frac{1}{\sqrt{2}} \left(\ket{\up}\otimes\ket{U} + \ket{\dn}\otimes\ket{D} \right) \,.
\end{equation}
For the second ball to move into a position consistent with measurement outcomes for the position of the first ball, according to Feynman, the gravitational field should possess an amplitude as well. And indeed, the gravitational potential according to semiclassical gravity would be that of half the mass at position $U$ and the other half at position $D$, \emph{regardless} of the measurement outcome for the position of the first ball.

This has been noticed and put to the test by Page and Geilker~\cite{pageIndirectEvidenceQuantum1981}. In their experiment, the link between the quantum states $\ket{\up}$ or $\ket{\dn}$ and the position of the first ball was as classical as it can be: it consisted of measuring the emission of $\gamma$ rays from a cobalt-60 source over 30 seconds, walking over to another table\footnote{Whether it was literally or only metaphorically another table is not clear from their paper.}, and setting the position of a torsion balance into one of two positions depending on the measured emission (above/below average). To nobody's surprise, the experiment confirms that whenever $\ket{\up}$ is measured the gravitational field is that of a ball at position $U$ and vice versa. This allows two possible conclusions:
\begin{enumerate}
    \item the gravitational field must, in fact, possess an amplitude, or
    \item the state of the system under consideration has \emph{not} been the entangled state~\eqref{eqn:pg-entangled}.
\end{enumerate}

That~\eqref{eqn:pg-entangled} cannot be the full story should, however, be old news to anyone who has ever encountered the quantum measurement problem~\cite{bellMeasurement1990,maudlinThreeMeasurementProblems1995} before. Only in a many worlds interpretation~\cite{everettRelativeStateFormulation1957} is the system expected to be in this state still after the spin measurement---with all the corresponding difficulties~\cite{kentManyworldsInterpretations1990} of such an interpretation. The consequence of the Page--Geilker experiment is that semiclassical gravity is incompatible with a no-collapse interpretation of QM.

\begin{important}{Conclusion 3.1}%
Semiclassical gravity, i.e.\ equation~\eqref{eqn:sce} for a single quantum field, is \emph{incomplete}. There must be a dynamical process, connected with the spin measurement, that leads to an objective reduction of the wave function:
\begin{align}\label{eqn:wf-collapse}
    \ket{\Psi} \to \begin{cases}
    \ket{\up}\otimes\ket{U} & \qquad\text{with probability 50\,\%} \\
    \ket{\dn}\otimes\ket{D} & \qquad\text{with probability 50\,\%\,.}
    \end{cases}
\end{align}
\end{important}

Page and Geilker are well aware of this possibility. They refute it, because a wave function collapse described by equation~\eqref{eqn:wf-collapse} seems to blatantly contradict the covariant conservation of energy-momentum, Wald's third axiom.

Interestingly, a similar argument could be made to refute QM, 
at least if defined in the traditional way~\cite{schrodingerGegenwartigeSituationQuantenmechanik1935}, including the postulate that the wave function after a measurement does not evolve from the wave function before measurement according to the \schr\ equation, but rather through projection on the corresponding eigenstate. In the same way, one could simply \emph{postulate} that the combined state $(g,\Psi)$ for metric and quantum field collapses upon `measurement', in violation of the semiclassical Einstein equations, and continues to evolve according to semiclassical gravity thereafter. Of course, one then faces the same measurement problem as in standard quantum theory, the crucial difference being that a many worlds or `operational' interpretation is not only difficult to reconcile with Born's rule but also in conflict with the observed reality.

If one does not take an agnostic point of view about measurement (as commonly accepted in non-gravitational quantum physics), semiclassical gravity requires the introduction of an objective collapse, with the instantaneous collapse~\eqref{eqn:wf-collapse} being only an effective, nonrelativistic description. We conclude:

\begin{important}{Conclusion 3.2}%
The objective reduction dynamics~\eqref{eqn:wf-collapse}, required to render semiclassical gravity a complete theory (Conclusion 3.1), is the nonrelativistic limit of a relativistic dynamical law for the fields compatible with the conservation law
\begin{equation}\label{eqn:sce-energy-momentum-conservation}
    \nabla^\mu \ev{\hat{T}_{\mu\nu}} = 0 \,.
\end{equation}
\end{important}

This leaves us with three options, \emph{none} of which can be excluded, as of yet:
\begin{enumerate}
    \item A consistent model for collapse, compatible with the conservation law~\eqref{eqn:sce-energy-momentum-conservation}, the Born rule probabilities~\eqref{eqn:wf-collapse} (or rather the generalisation to arbitrary states), as well as all other observations in quantum theory, especially the violation of Bell's inequalities, is fundamentally \emph{impossible}.
    \item There is a \emph{new} process, to be modelled outside the formalism for quantum fields on curved spacetime with backreaction, that provides an explanation of the collapse~\eqref{eqn:wf-collapse} compatible with~\eqref{eqn:sce-energy-momentum-conservation}.
    \item A consistent explanation for the \emph{effective} collapse~\eqref{eqn:wf-collapse} can be given \emph{within} the theory of semiclassical gravity (by taking into account \emph{all} matter fields and their interactions).
\end{enumerate}
Clearly, the first possibility is the perspective taken by Page and Geilker, among many others, and would necessitate \emph{some} sort of quantisation of GR if it were true. The second possibility includes relativistic generalisations of collapse models~\cite{bassiModelsWavefunctionCollapse2013}. The last certainly constitutes the most interesting alternative, that the explanation of wave function collapse could somehow lie within semiclassical gravity itself, although it is also the most speculative one.

\subsubsection{The role of the density operator and its dynamics}
The quantum mechanical measurement postulate~\eqref{eqn:wf-collapse} predicts stochastic outcomes. Semiclassical gravity by itself, on the contrary, is a deterministic model---as are QM and \QFT\ without the collapse postulate. On the other hand, even in classical, deterministic theories stochastic phenomena are a regular occurrence in situations of many degrees of freedoms with incomplete information about the precise initial conditions.

The quantum mechanical formalism deals with these twofold statistics with the introduction of the density operator: given an ensemble $\{(p_j,\ket*{\psi_j})\}_j$ of pure Hilbert space states $\ket*{\psi_j}$ that are expected with classical probabilities $p_j$, the density operator of the system is given by
\begin{equation}\label{eqn:density-operator}
    \hat{\varrho} = \sum_j p_j \ket{\psi_j}\bra{\psi_j} \,.
\end{equation}
Two ensembles are called \emph{equivalent} if they have the same density operator. The probability for any outcome $o$ of a projective measurement of an operator $\hat{O}$ is given by the trace $\tr \ket{o}\bra{o}\hat{\varrho}$ of the projector on the corresponding eigenstate\footnote{We assume, for simplicity, that operators have non-degenerate spectra, although the generalisation to degenerate eigenvalues is straightforward.} $\ket{o}$ to $o$ and the density operator. The information accessible via projective measurements is, therefore, entirely encoded in $\hat{\varrho}$.

The density operator serves a second role in conventional QM. Pure states in a composite Hilbert space, $\ket{\Psi} \in \mathcal{H}_1 \otimes \mathcal{H}_2$, are generally entangled (non-separable). There is no pure state $\ket*{\psi_1} \in \mathcal{H}_1$ from which one could obtain probabilities for measurement outcomes of an operator $\hat{O}_1 \in \mathrm{End}(\mathcal{H}_1)$. Instead, if $\hat{\varrho} = \ket{\Psi}\bra{\Psi}$ is the (pure state) density operator, the probabilities for $\hat{O}_1$ can be derived from the partial trace $\hat{\varrho}_1 = \tr_{\mathcal{H}_2} \hat{\varrho}$ over the second Hilbert space.

For a linear Hamiltonian, the time evolution of the Hilbert space states $\ket*{\psi_j}$ in equation~\eqref{eqn:density-operator} induces the time evolution law $\I \hbar \partial_t \hat{\varrho} = [\hat{H},\hat{\varrho}]$ for the density operator. It has a closed form, implying that equivalence of ensembles is a property preserved under time evolution. Even the partial trace $\hat{\varrho}_1$ for a subsystem obeys a closed time evolution law---in the case of a Markovian dynamics a master equation in Lindblad form $\hbar \partial_t \hat{\varrho} = -\I [\hat{H},\hat{\varrho}] + \hbar \mathcal{L}(\hat{\varrho})$---which is linear in $\hat{\varrho}$ (although not generally unitary).

Considering, instead, the nonlinear evolution law~\eqref{eqn:sne-single-part}, the spatial density matrix $\varrho_t(\vec x,\vec x') = \sum_j p_j \psi_j(t,\vec x)\psi_j^*(t,\vec x')$ evolves according to
\begin{align}\label{eqn:master-eq-sn}
    \partial_t \varrho_t(\vec x,\vec x') 
    &= -\frac{\I}{\hbar} [\hat{H}_0,\varrho_t(\vec x,\vec x')] 
    + \sum_j \frac{\I\,G\,m^2}{\hbar} p_j \psi_j(t,\vec x)\psi_j^*(t,\vec x') \nnl
    &\bleq \times
    \int \D^3 x'' \abs{\psi_j(t,\vec x'')}^2 \left(
    \frac{1}{\abs{\vec x - \vec x''}} - \frac{1}{\abs{\vec x' - \vec x''}}\right) \,.
\end{align}
It does not have a closed form and, therefore, does not preserve equivalence of ensembles. Although it is possible to calculate a density operator \emph{at any given time} in order to obtain probabilistic predictions, the dynamics cannot be described in terms of the density operator for nonlinear systems.

\begin{example}{Example}%
Let $\ket*{\psi_{1,2}}$ be stationary solutions of the \schr--Newton equation~\eqref{eqn:sne-single-part} centred around $\vec x_{1,2}$, respectively, and define $\ket*{\psi_\pm} = \frac{1}{\sqrt{2}} \left(\ket*{\psi_1} \pm \ket*{\psi_2}\right)$; note that those superposition states are time dependent. The ensemble $A = \{(\tfrac12,\ket*{\psi_1}),(\tfrac12,\ket*{\psi_2})\}$ then has the constant density matrix $\varrho_A(\vec x,\vec x')$. The ensemble $B = \{(\tfrac12,\ket*{\psi_+}),(\tfrac12,\ket*{\psi_-})\}$ has the same initial density matrix. Nonetheless, its density matrix evolves in time.

The stationary wave functions $\psi_j(\vec x)$ are real valued and assumed to be sharply peaked around $\vec x_j$. The initial wave functions $\psi_\pm(0,\vec x)$ then are real valued, as well, and in absence of an external potential, $V_\text{ext} = 0$, one finds
\begin{align}
    \psi_\pm(t,\vec x) &= \exp\left[-\frac{\I t}{\hbar} \left(\frac{\hat{\vec p}^2}{2m} 
    - G m^2 \int \D^3 x' \frac{\abs{\psi_\pm(0,\vec x')}^2}{\abs{\vec x - \vec x'}} \right)\right] \psi_\pm(0,\vec x) \nnl
    &\approx \exp\left[-\I t \left(\mathcal{L}_1 + \mathcal{L}_2 \right)\right] \psi_\pm(0,\vec x) \nnl
    &\approx \left[1 - \I t \left(\mathcal{L}_1 + \mathcal{L}_2 \right)
    - \frac{t^2}{2} \left(\mathcal{L}_1^2 + \mathcal{L}_2^2 + \{\mathcal{L}_1,\mathcal{L}_2\} \right)\right] \psi_\pm(0,\vec x) \,,
    \intertext{where we approximated the $\psi_j$ as sharply peaked in the first step, and used the Zassenhaus formula expanding to quadratic order in time in the second, defining}
    \mathcal{L}_j &= \frac{\hat{\vec p}^2}{4 \hbar m} 
    - \frac{G m^2 }{2 \hbar \abs{\vec x - \vec x_j}} \,.
\end{align}
This approximates the nonlinear evolution by the application of a linear operator. Similarly, the stationarity of the $\psi_j(\vec x)$ implies they must be close to states in the kernel of the operators $1 - \exp(-\I t \mathcal{L}_j)$. Linearising in $\vec x$ and ignoring constant phase contributions, we have
\begin{align}
     \psi_\pm(t,\vec x) &\approx \psi_\pm(0,\vec x) 
     - \frac{\I t}{\sqrt{2}} \left(\mathcal{L}_2 \psi_1(\vec x) \pm\mathcal{L}_1 \psi_2(\vec x)\right) \nnl
     &\bleq -\frac{t^2}{2\sqrt{2}} \left(\mathcal{L}_2^2 \psi_1(\vec x) \pm \mathcal{L}_1^2 \psi_2(\vec x) 
     + [\mathcal{L}_1,\mathcal{L}_2] \left(\psi_1(\vec x) \mp \psi_2(\vec x)\right) \right) \,,
\end{align}
and thus the difference in the diagonal elements of the density matrix is
\begin{align}
    \Delta P(\vec x) &= \varrho_B(\vec x,\vec x) - \varrho_A(\vec x,\vec x) \nnl 
    &\approx
    \frac{t^2}{2} \Big( (\mathcal{L}_2 \psi_1(\vec x))^2 - \psi_1(\vec x) \mathcal{L}_2^2 \psi_1(\vec x)
    - \psi_1(\vec x) [\mathcal{L}_1, \mathcal{L}_2] \psi_1(\vec x)
    \nnl &\bleq 
    + (\mathcal{L}_1 \psi_2(\vec x))^2 - \psi_2(\vec x) \mathcal{L}_1^2 \psi_2(\vec x)
    - \psi_2(\vec x) [\mathcal{L}_2, \mathcal{L}_1] \psi_2(\vec x)
    \Big)  \,.
\end{align}
It is tedious and not particularly insightful to attempt to evaluate these expressions explicitly. Nonetheless, we can get a good idea of the dynamics, considering that the operators $\mathcal{L}_1$ acting on $\psi_2$ and vice versa induce a motion of the peaks comparable to the gravitational attraction between two particles, with both the gravitational mass and the kinetic energy split equally between the two peaks of the superposition.
\end{example}

\subsection{Causality of semiclassical gravity}
\label{subsubsec:causality}
The property discussed in the previous subsection, that equivalent ensembles evolve into distinguishable ones, has consequences for the possibility to send signals faster than light. Although the possibility of faster-than-light signalling has been discussed before~\cite{eppleyNecessityQuantizingGravitational1977}, the argument is often attributed to Gisin~\cite{gisinStochasticQuantumDynamics1989} who introduces the following lemma\footnote{See appendix B of the preprint version of reference~\cite{bassiNofasterthanlightsignalingImpliesLinear2015} for a complete proof.}:
\begin{lemma}
Let $\{(p_j,\ket*{\psi_j})\}_{j = 1,\dots,n}$ and $\{(q_k,\ket*{\chi_k})\}_{k = 1,\dots,m}$ be two equivalent ensembles of states $\ket{\psi_j}$, $\ket{\chi_k} \in \mathcal{H}$. Let $\mathcal{K}$ be a complex Hilbert space of dimension $l \geq \max(n,m)$ and $\{\ket*{\alpha_i}\}_{i=1,\dots,l}$, $\{\ket*{\beta_i}\}_{i=1,\dots,l}$ two orthonormal bases of $\mathcal{K}$.
Then there is a state vector $\ket{\Psi} \in \mathcal{H} \otimes \mathcal{K}$ such that
\begin{equation}\label{eqn:gisin-state}
    \ket{\Psi}
    = \sum_{j=1}^n \sqrt{p_j} \ket{\psi_j} \otimes \ket{\alpha_j}
    = \sum_{k=1}^m \sqrt{q_k} \ket{\chi_k} \otimes \ket{\beta_k} \,.
\end{equation}
\end{lemma}
The state $\ket{\Psi}$ has the reduced density matrix
\begin{equation}\label{eqn:red-density-harvey}
    \hat{\varrho}_\mathcal{H} = \sum_{j=1}^n p_j \ket{\psi_j}\bra{\psi_j}
    = \sum_{k=1}^m q_k \ket{\chi_k}\bra{\chi_k}
\end{equation}
when traced over the Hilbert space $\mathcal{K}$. Enter Harvey and Krista, both adept experimenters stationed in remote locations. At an earlier time, the state $\ket{\Psi}$ has been prepared and distributed such that Harvey is able to perform local measurements with respect to operators in $\mathcal{H}$. Krista, on the other hand, can decide to perform a measurement with respect to one of the self-adjoint operators $\hat{A}, \hat{B} \in \mathrm{End}(\mathcal{K})$, whose matrices are diagonal relative to the bases $\{\ket{\alpha_i}\}_i$ and $\{\ket{\beta_i}\}_i$, respectively. According to the quantum mechanical measurement postulate, the global state after Krista's measurement is one of the projections
\begin{equation}\label{eqn:proj-state-alpha}
    \ket{\Psi} \to \frac{\ket{\alpha_i}\bra{\alpha_i}\ket{\Psi}}{\abs{\braket{\alpha_i}{\Psi}}}
    = \ket{\psi_i} \otimes \ket{\alpha_i} \qq{with probability} p_i
\end{equation}
if she chooses to measure with respect to $\hat{A}$, and one of the projections 
\begin{equation}\label{eqn:proj-state-beta}
    \ket{\Psi} \to \frac{\ket{\beta_i}\bra{\beta_i}\ket{\Psi}}{\abs{\braket{\beta_i}{\Psi}}}
    = \ket{\chi_i} \otimes \ket{\beta_i} \qq{with probability} q_i
\end{equation}
if she chooses to measure with respect to $\hat{B}$. Due to the separability of the collapsed states \eqref{eqn:proj-state-alpha} and \eqref{eqn:proj-state-beta}, and without knowledge about Krista's measurement outcome, Harvey's subsystem is then represented by the ensembles $\{(p_j,\ket*{\psi_j})\}_{j = 1,\dots,n}$ and $\{(q_k,\ket*{\chi_k})\}_{k = 1,\dots,m}$, respectively. Being equivalent ensembles, these are described by the same density operator~\eqref{eqn:red-density-harvey}.

The crucial ascertainment is now that, because with a linear dynamical law equivalent ensembles remain equivalent, Harvey has no means to detect which ensemble he is dealing with. The probabilities of all possible measurements he can perform on the Hilbert space $\mathcal{H}$ are fully determined by $\hat{\varrho}_\mathcal{H}(t)$ at any time $t$. With a nonlinear evolution law, on the other hand, the two ensembles evolve differently, as discussed in the previous subsection for the example of the \schr--Newton equation. After some finite time $t$ they become distinguishable. If Krista and Harvey previously agreed to encode a binary signal with Krista's choice of basis, then this signal reaches Harvey before any light signal could, provided the distance between them is larger than $ct$.

Before jumping to the conclusion that this sort of faster-than-light signalling is detrimental to semiclassical gravity, one may at least consider a series of potential loopholes in the argument:
\begin{enumerate}
    \item The fact that a state $\ket{\Psi} \in \mathcal{H} \otimes \mathcal{K}$ with the property~\eqref{eqn:gisin-state} exists does not necessarily imply that this state can ever be created as the consequence of some dynamical law, given the initial conditions of the universe. There is at least a theoretical possibility that the dynamics are such that they prevent \emph{physical} states from ever approaching such a state. 
    \item Standard quantum (field) theory permits only local interactions. Hence, the entangled state $\ket{\Psi}$ must ultimately be created locally (either directly or via some third system) and brought to Krista's and Harvey's locations at separation $d \leq c t_0$ in final time $t_0$. The nonlinear evolution that results in the distinguishability of the initially equivalent ensembles, therefore, already acts during the time interval $[-t_0,0]$ of separation, and not only during the time interval $[0,t]$ after Krista's measurement. The initial state $\ket{\Psi(t=0)}$ is then \emph{itself} the result of the nonlinear evolution and cannot simply be assumed to be of the form~\eqref{eqn:gisin-state}.
    \item Projective measurement with an instantaneous collapse as in equations~\eqref{eqn:proj-state-alpha} and \eqref{eqn:proj-state-beta} is a frame dependent assumption, already in obvious contradiction with relativity. Without a Lorentz invariant formulation of the collapse, one could justifiably ask why one should even bother about the possibility of faster-than-light signalling in a nonrelativistic model.
    \item Even if one accepts the possibility of faster-than-light signalling, notwithstanding the three previous points, it is not evident that this specific form of signalling would result in a conflict with causality. Such a conflict would occur if Harvey used a second entangled system to signal back to Krista \emph{and} that second signal would arrive at Krista's location \emph{before} she makes her choice of basis. This procedure requires \emph{two} measurement processes which must be instantaneous in \emph{different} frames of reference. Hence, a complete thought experiment for causality violation goes beyond the scenario considered above.
\end{enumerate}
Regarding the third point, specifically, and taking into account the requirement for an objective collapse of the wave function established above, one would like to consider a scenario for faster-than-light signalling including the collapse dynamics. Relativistic models for the wave function collapse are still only sparsely developed~\cite{bedinghamCollapseModelsRelativity2020}. 

Penrose~\cite{penroseGravityRoleQuantum1996,penroseQuantumComputationEntanglement1998} suggests that collapse should occur in such a way that a spatial superposition of two classical states decays into a classical state with a rate proportional to the gravitational self-energy between the mass distributions belonging to the two states in superposition. Di\'osi's nonrelativistic collapse model~\cite{diosiModelsUniversalReduction1989} implements this idea. It also requires the introduction of a length scale $r_c$, acting as a cutoff in order to prevent divergences. Conforming with other nonrelativistic collapse models, it is based on a stochastic evolution law for Hilbert space states and a \emph{linear} evolution of the density operator.

If we maintain the idea of a cutoff length scale and only focus on superposition states of two wave functions $\psi_{1,2}$ narrowly peaked around $\vec x_{1,2}$, respectively, we can give a more ad hoc description of collapse:
\begin{svgraybox}
Whenever a superposition $\psi(\vec x) = \alpha \psi_1(\vec x) + \beta \psi_2(\vec x)$ of two narrowly peaked wave functions $\psi_1(\vec x) \sim \delta^{(3)}(\vec x - \vec x_1)$ and $\psi_2(\vec x) \sim \delta^{(3)}(\vec x - \vec x_2)$ for a particle of mass $m$ exceeds the cutoff scale, $\abs{\vec x_1 - \vec x_2} \geq r_c$, the state undergoes a \emph{collapse}
\begin{equation}
    \psi(\vec x) \to \begin{cases}
    \psi_1(\vec x) & \text{with probability } \abs{\alpha}^2 \\
    \psi_2(\vec x) & \text{with probability } \abs{\beta}^2
    \end{cases}
\end{equation}
within the time scale
\begin{equation}
    \tau_c = \frac{\hbar \, r_c}{G \, m^2} \,.
\end{equation}
\end{svgraybox}
This prototype of a collapse dynamics remains agnostic about the dynamics of more complex states that are not simple superpositions of two localised peaks of a single particle wave function. 

For a spatial separation below $r_c$, as well as in addition to the collapse dynamics for larger separations, an evolution according to the \schr--Newton equation~\eqref{eqn:sne-single-part} is assumed. It predicts that the two peaks, initially at $\vec x_{1,2}$, shift towards
\begin{equation}
    \widetilde{\vec x}_1 = \vec x_1 - \frac{\vec x_1 - \vec x_2}{\abs{\vec x_1 - \vec x_2}} \abs{\alpha}^2 \delta x \,,\qquad
    \widetilde{\vec x}_2 = \vec x_2 + \frac{\vec x_1 - \vec x_2}{\abs{\vec x_1 - \vec x_2}} \abs{\beta}^2 \delta x \,,
\end{equation}
decreasing their initial distance $\Delta x = \abs{\vec x_1 - \vec x_2}$ by
\begin{equation}\label{eqn:causality-delta-x}
    \delta x \approx \frac{G m t^2}{2 \Delta x^2} \,,
\end{equation}
where we assume $\delta x \ll \Delta x$.
In order to resolve this decrease against the expected value $\delta x = 0$ for a classical mixture of the states $\psi_{1,2}$ with probabilities $\abs{\alpha}^2$ and $\abs{\beta}^2$, it must be larger than the free spreading of the wave function. Due to the position-momentum uncertainty relation, this spreading for a state with initial position uncertainty $\delta \xi$ and momentum uncertainty $\delta p$ is limited by
\begin{equation}
    \delta \xi + \frac{t}{m} \delta p \geq \delta \xi + \frac{\hbar t}{2 m \, \delta \xi}
    = \sqrt{\frac{\hbar t}{2 m}} \left(\zeta + \frac{1}{\zeta}\right)
    \geq \sqrt{\frac{2 \hbar t}{m}} \,,
\end{equation}
and hence the minimal resolution after time $t$ is
\begin{equation}\label{eqn:causality-m3}
    \delta x^4 \geq \frac{4 \hbar^2 t^2}{m^2} 
    \stackrel{\eqref{eqn:causality-delta-x}}{=} \frac{8 \hbar^2 \, \delta x \, \Delta x^2}{G m^3}
    \qquad \Rightarrow \qquad
    m^3 \geq \frac{8 \hbar^2 \, \Delta x^2}{G \, \delta x^3}
    \gg \frac{8 \hbar^2}{G \, \delta x} \,.
\end{equation}
Therefore, a minimum mass is required in order to achieve the necessary resolution. However, if we account for a dynamical collapse of the wave function according to the above ad hoc description, a larger mass implies a faster collapse and we must take care that the superposition is maintained throughout the entire time $t$ of the experiment by requiring $t < \tau_c$. One then has
\begin{equation}\label{eqn:causality-rc}
    r_c = \frac{G \, m^2 \, \tau_c}{\hbar}
    > \frac{G \, m^2 \, t}{\hbar}
    \stackrel{\eqref{eqn:causality-delta-x}}{=}
    \sqrt{\frac{2 G \, m^3 \, \Delta x^2 \, \delta x}{\hbar^2}}
    \stackrel{\eqref{eqn:causality-m3}}{\geq}
    \frac{4 \, \Delta x^2}{\delta x}
    \gg \Delta x \gg \delta x \,.
\end{equation}
Any shift $\delta x$ above the length scale $r_c$ would always remain unobservable, because the collapse happens too fast. Combining equations~\eqref{eqn:causality-rc} and \eqref{eqn:causality-m3} yields
\begin{equation}
    r_c \geq \frac{32 \hbar^2}{G m^3} \left(\frac{\Delta x}{\delta x}\right)^4
    \gg \frac{32 \hbar^2}{G m^3} \,.
\end{equation}
This limit only holds for separations above the size of the particle, $\Delta x > 2R$. For a spherical particle with homogeneous mass density one then finds
\begin{equation}
    m = \frac{4 \pi}{3} \rho \, R^3 < \frac{\pi}{6} \rho \, \Delta x^3
    \qquad \Rightarrow \qquad
    r_c \gg \left(\frac{6912 \,\hbar^2}{\pi^3 \, G \, \rho^3}\right)^{1/10} \,,
\end{equation}
which implies that any value for $r_c$ below \unit{140}{\nano\meter} makes it impossible to resolve the required separation even for the densest elements.

For separations $\Delta x < 2 R$ one finds instead of equation~\eqref{eqn:causality-delta-x} that
\begin{equation}
    \delta x \approx \frac{2 \pi}{3} G \, \rho \, \Delta x \, t^2 \left( 1 + \order{\frac{\Delta x}{R}} \right) \,.
\end{equation}
Inserting this into equation~\eqref{eqn:causality-m3} yields
\begin{equation}\label{eqn:causality-m3b}
    \delta x^4 \geq \frac{4 \hbar^2 t^2}{m^2} 
    = \frac{6 \, \hbar^2 \, \delta x}{\pi \, G \, \rho \, \Delta x \, m^2} 
    \qquad \Rightarrow \qquad
    \delta x \geq \left(\frac{6 \, \hbar^2}{\pi \, G \, \rho \, \Delta x \, m^2}\right)^{1/3}
\end{equation}
and instead of equation~\eqref{eqn:causality-rc} one finds, using $2 R > \Delta x > \delta x$ and \eqref{eqn:causality-m3b},
\begin{equation}
    r_c > \sqrt{\frac{3\, G \,m^4\, \delta x}{2 \pi\, \rho\, \hbar^2 \,\Delta x}}
    > R \,.
\end{equation}
The radius is limited by the condition~\eqref{eqn:causality-m3b}, which with $\Delta x < 2 R$ yields
\begin{equation}
    R^3 > \frac{\delta x^3}{8} \geq \frac{3 \, \hbar^2}{4 \pi \, G \, \rho \, \Delta x \, m^2}
    = \frac{27 \, \hbar^2}{64 \pi^3 \, G \, \rho^3 \, \Delta x \, R^6}
    > \frac{27 \, \hbar^2}{128 \pi^3 \, G \, \rho^3 \, R^7}
\end{equation}
implying
\begin{equation}
    r_c > \left(\frac{27 \, \hbar^2}{128 \pi^3 \, G \, \rho^3}\right)^{1/10}
    \gtrsim \unit{50}{\nano\meter}\,.
\end{equation}

These minimum values required for $r_c$ are orders of magnitude above the parameter range usually assumed in collapse models, as well as above parameters excluded by observation. For instance, levitated nanoparticles~\cite{delicCoolingLevitatedNanoparticle2020} only pose a limit at about $r_c \gtrsim \unit{1}{\femto\meter}$. Even the recent underground tests~\cite{donadiUndergroundTestGravityrelated2021}, which consider radiation emission rather than spatial superposition states and are of limited use for constraining the prototype collapse dynamics used here, only restrict the cutoff parameter of the Di\'osi model to $r_c \gtrsim \unit{500}{\pico\meter}$. If there is a consistent relativistic description of collapse that approximates the prototypical model here for nonrelativistic superpositions of two sharp peaks, there is a large parameter range for which it would effectively prevent at least the most simple ideas to use the nonlinearity of the \schr--Newton equation for faster-than-light signalling, while being perfectly consistent with observation.

\subsection{Other schemes to include backreaction for quantum matter on a classical spacetime}
So far, we have only discussed the specific model for semiclassical gravity described by the semiclassical Einstein equations~\eqref{eqn:sce}. The definition of semiclassical gravity given at the beginning of section~\ref{sec:qm-backreaction} also allows for other ways to introduce quantum matter as the source of curvature in Einstein's equations.

One alternative presents itself in the context of collapse models~\cite{bassiModelsWavefunctionCollapse2013}. In these models, the Fock space state obeys a stochastic differential equation, e.g.,
\begin{align}
    \ket{\D \psi}_t &= \Big[-\frac{\I}{\hbar} \hat{H} \,\D t
    + \sqrt{\gamma} \int \D^3 x \, \left(\hat{\rho}_\text{reg}(\vec x) - \expval{\hat{\rho}_\text{reg}(\vec x)}_t\right) \,\D W(t,\vec x)
    \nnl &\bleq
    - \frac{\gamma}{2} \int \D^3 x \,\left(\hat{\rho}_\text{reg}(\vec x) - \expval{\hat{\rho}_\text{reg}(\vec x)}_t\right)^2 \,\D t \Big] \ket{\psi}_t
\end{align}
in the continuous spontaneous localisation (CSL) model, where $\hat{H}$ is the usual Hamiltonian, $\hat{\rho}_\text{reg}$ the regularised mass density operator, $W(t,\vec x)$ describes an ensemble of independent stochastic Wiener processes (one for every point $\vec x$), and $\gamma$ is a free coupling parameter. A second free parameter comes from the regularisation length scale $r_c$ for the mass density $\hat{\rho}_\text{reg}$. The density operator satisfies the stochastic master equation
\begin{align}\label{eqn:sme}
    \D \hat{\varrho}(t) &= -\frac{\I}{\hbar} \comm{\hat{H}}{\hat{\varrho}(t)} \,\D t
    -\frac{\gamma}{8 \hbar^2}
    \comm{\hat{\rho}_\text{reg}(\vec x)}{\comm{\hat{\rho}_\text{reg}(\vec x)}{\hat{\varrho}(t)}} \,\D t
    \nnl &\bleq + \frac{\gamma}{2 \hbar^2}
    \acomm{\hat{\rho}_\text{reg}(\vec x) - \expval{\hat{\rho}_\text{reg}(\vec x)}}{\hat{\varrho}(t)} \, \D W(t,\vec x) \,.
\end{align}
The characteristic feature of collapse models is that after averaging over the noise term in the second line of~\eqref{eqn:sme}, the density operator satisfies the linear Gorini--Kossakowski--Sudarshan--Lindblad equation~\cite{goriniCompletelyPositiveDynamical1976,lindbladGeneratorsQuantumDynamical1976} (i.e., a master equation in Lindblad form), which ensures that equivalent ensembles evolve into equivalent ensembles and no faster-than-light signalling can occur, as detailed in the previous subsection.

An intuitive way to couple the mass density to gravity, as pointed out by Tilloy and Di\'osi~\cite{tilloySourcingSemiclassicalGravity2016}, is then to use the \emph{signal}
\begin{equation}\label{eqn:signal}
    \rho(t,\vec x) = \expval{\hat{\rho}_\text{reg}(\vec x)}_t + \delta \rho_t \,,
\end{equation}
where $\delta \rho_t$ are the noise fluctuations resulting from the stochastic part in the second line of equation~\eqref{eqn:sme}. One obtains a classical Newtonian potential
\begin{equation}
    \Phi(t,\vec x) = -G\,\int \D^3 x' \, \frac{\rho(t,\vec x')}{\abs{\vec x - \vec x'}} \,,
\end{equation}
entering into the \schr\ equation instead of the potential~\eqref{eqn:sc-newton-pot}. In the language of standard quantum physics, \eqref{eqn:signal} can be interpreted as the information retrievable via weak measurement of the mass density, which is then fed back into the dynamics as the source of the gravitational field. The joint dynamics contain decoherence terms both from the stochastic noise and from the gravitational potential~\cite{kafriClassicalChannelModel2014,tilloySourcingSemiclassicalGravity2016}. In the language of collapse models, the gravitational field is sourced by the collapse events (flashes) in spacetime~\cite{tilloyGhirardiRiminiWeberModelMassive2018}.

Related is the concept of hybrid classical-quantum dynamics~\cite{hallInteractingClassicalQuantum2005}. Let $\mathcal{H}$ be a Hilbert space of some quantum system and $\mathcal{M} \cong \mathbb{R}^{2n}$ a classical phase space. Then a \emph{hybrid state} is given by
\begin{equation}
    \hat{\varrho}_{cq} : \mathcal{M} \to \mathrm{End}(\mathcal{H}) \qquad\text{with}\qquad
    \int_\mathcal{M} \D z \, \tr \hat{\varrho}_{cq}(z) = 1 \,,
\end{equation}
i.e., by assigning a subnormalised density matrix to every point in phase space. It can be understood as the product $\hat{\varrho}_{cq} = \int \D z \, \varrho_c(z) \ket{z}\bra{z} \otimes \hat{\varrho}_q(z)$ of a density matrix $\hat{\varrho}_q$ for the quantum system and a probability distribution $\varrho_c(z)$ of the orthonormal classical phase space states $\ket{z}$. This formalism has been used by Albers et al.~\cite{albersMeasurementAnalysisQuantum2008} in order to couple a scalar quantum field to classical scalar gravity. Oppenheim~\cite{oppenheimPostquantumTheoryClassical2021} recently proposed a model based on the ADM formalism~\cite{arnowittDynamicalStructureDefinition1959} of GR, which describes a Hamiltonian evolution of 3-manifolds with a time parameter $t$. The classical phase space $\mathcal{M}$ is formed by the 3-metrics $g$ and their canonical momenta $\pi$. The Hamiltonian and momentum constraints $H$ and $P^i$ in the ADM Lagrangian,
\begin{equation}
    \mathcal{L} = -g_{ij} \partial_t \pi^{ij} - N H - N_i P^i - 2 \partial_i \left( \pi^{ij} N_j - \frac{1}{2} \pi N^i + \nabla^i N \sqrt{g} \right)\,,
\end{equation}
where $N$, $N_i$ are the lapse and shift function, respectively, $\pi = g_{ij} \pi^{ij}$ denotes the trace, $g$ the metric determinant, and the covariant derivative $\nabla$ is taken with respect to the 3-metric, are then replaced by corresponding operators,
\begin{align}
    H &= -\sqrt{g} \left[R + g^{-1} \left(\frac{1}{2} \pi^2 - \pi^{ij} \pi_{ij} \right) \right] \hat{\mathrm{I}}
    + f^{\alpha\beta} \hat{L}^\dagger_\alpha \hat{L}_\beta\\
    P^i &= -2 \nabla_j \pi^{ij} \,\hat{\mathrm{I}} + g_i^{\alpha\beta} \hat{L}^\dagger_\alpha \hat{L}_\beta \,,
\end{align}
acting on hybrid states, with $\hat{\mathrm{I}}$ the identity, $\hat{L}_\alpha$ a set of Lindblad operators, and $f^{\alpha\beta}$, $g_i^{\alpha\beta}$ coefficient functions on 3-space which depend on the classical state $(g,\pi)$. The scalar curvature $R$ is, again, to be taken with respect to the 3-metric $g$.

This formalism allows for a consistent semiclassical theory without the need to introduce any explicit assumptions about wave function collapse. It is, however, a probabilistic theory of \emph{statistical ensembles} of 3-metrics rather than a deterministic theory for a single spacetime. The large freedom of choice for the proper Lindblad operators and coefficient functions makes it currently difficult to arrive at experimental predictions.

Finally, Kent~\cite{kentNonlinearitySuperluminality2005,kentTestingCausalQuantum2018} has developed a framework of causal quantum theory. Consider a time orientable, globally hyperbolic Lorentzian 4-manifold $(\mathcal{M},g)$ with a foliation $\mathcal{M} = \cup_{t \in \mathbb{R}} \mathcal{S}_t$ of spacelike hypersurfaces $\mathcal{S}_t$, that allows to define a unitary evolution law $\psi(t,\vec x) = U(t,t_0) \psi(t_0,\vec x)$ with the initial wave function\footnote{More generally, one can consider a quantum field operator evolving between Cauchy surfaces according to the Tomonaga--Schwinger equation.} defined on $\mathcal{S}_{t_0}$. Further assume a set of local measurement events $M_i = (t_i,\vec x_i; \hat{O}_i,\ket{o_i})$ ($i=1,\dots,n$) of eigenstates $\ket{o_i}$ of operators $\hat{O}_i$ at spacetime points $(t_i,\vec x_i \in \mathcal{S}_{t_i})\in \mathcal{M}$ which are time-ordered, $t_0 < t_1 < t_2 < \dots < t_n < t$. In standard quantum theory, the final state is obtained as the unitary evolution with $U(t,t_0)$ conditioned over the measurement outcomes:
\begin{equation}\label{eqn:cqt-time-evolution}
    \psi(t,\vec x) = U(t,t_n) \left(\prod_{i=n,n-1,\dots,1} \ket{o_i}\bra{o_i} U(t_i,t_{i-1})\right)
    \psi(t_0,\vec x) \,.
\end{equation}
In the causal theory proposed by Kent, for any point $\vec x \in \mathcal{S}_t$ the time evolution law~\eqref{eqn:cqt-time-evolution} holds, however with the important distinction that only those measurement events $M_i$ with $(t_i,\vec x_i)$ in the \emph{past light cone} of $(t,\vec x)$ are conditioned over. This results in a definition of \emph{local states} and predictions that differ from standard quantum theory, although thus far not being in any obvious contradiction with observation.

A nonlinear modification of the dynamics based on these \emph{local} states does not result in the problem of faster-than-light signalling discussed previously. One could, therefore, attempt to formulate a semiclassical theory of gravity in which spacetime curvature is sourced by the local states, in order to avoid issues with causality. To date, such a model has not been developed.

\section{Concluding remarks}

In the winter semester 1923/24, Max Born delivered
a lecture on the Bohr--Sommerfeld quantisation 
in atomic physics which he subsequently published 
as a book entitled `Atommechanik' \cite{bornAtommechanik1925}. In that book 
he laid out with almost axiomatic precision the 
known formal principles from classical physics, 
like the Hamilton--Jacobi theory and its application 
to perturbation theory (which had proved very 
successful in astronomy), and their application 
to the Bohr--Sommerfeld quantisation. At that time 
it was clear to everyone in the field that this 
loose collection of rather ad-hoc 
`quantisation rules' would eventually be replaced 
by something with a proper logical foundation, a 
real \emph{theory}. So why did Born put all his 
efforts into that book project, given the premature 
state of a real understanding? His own answer is 
this: \emph{`... we are attempting a deductive 
presentation of atomic theory. The reservations, that 
the theory is not sufficiently developed, I wish to
disperse with the remark that we are dealing with 
a test case, a logical experiment, the meaning of 
which just lies in the determination of the limits 
to which the principles of atomic and quantum 
physics succeed, and to pave the ways which shall 
lead us beyond those limits.'}

Our contribution to 
this collection should likewise be regarded as
a `logical experiment', the meaning of which lies in 
in the determination of the limits to which 
classical or semi-classical gravity can be combined
with quantum mechanics. In the same vein, we believe that from 
these considerations we will receive useful hints as to where we 
really need to go beyond those limits. 

On these grounds, we propose a pragmatic attitude which puts first things first: From an experimental point
of view, the interface of classical gravity and quantum mechanics on the one hand as well as semiclassical gravity on the other hand present promising opportunities. Systematic post-Newtonian descriptions of the coupling between quantum matter and gravity make reliable testable predictions as to the influence of the gravitational field on the dynamics of quantum systems. Likewise, both the \schr--Newton
equation and models for objective wave function
collapse offer concrete experimental possibilities. 
Experiments of this type require comparably small
improvements over existing technology, and should be fully explored, with the
hope that they yield effects that can guide us towards the correct fundamental
theory. From the theoretical perspective, one should take conceptual challenges
within the established fundamental theories---such as the ones presented by
us---seriously in order to understand the precise breaking points of GR and
RQFT. A more thorough understanding of where \emph{exactly} they fail (and to which extent) when
considered jointly is expected to reveal valuable information for the search of a fully
consistent theory of gravitating quantum matter.

%%%%%%%%%%%%%%%%%%%%%%%%%%%%%%%%%%%%%%%%%%%%%%%%%%%%%%%%%%%%%%%%%%%%%%
%%%%%%%%%%%%%%%%%%%%%%%%%%%%%%%%%%%%%%%%%%%%%%%%%%%%%%%%%%%%%%%%%%%%%%

%%%%%%%%%%%%%%%%%%%%%%%%%%%%%%%%%%%%%%%%%%%%%%%%%%%%%%%%%%%%%%%%%%%%%%
%%%%%%%%%%%%%%%%%%%%%%%%%%%%%%%%%%%%%%%%%%%%%%%%%%%%%%%%%%%%%%%%%%%%%%

\begin{acknowledgement}
A.\,G.\ acknowledges funding from the VolkswagenStiftung. D.\,G.\ and P.\,K.\,S.\ acknowledge support from the Deutsche Forschungsgemeinschaft via the Collaborative Research Centre 1227 `DQ-mat', projects A05 and B08, at Leibniz University Hannover. All three authors would like to thank the organisers of the 740\textsuperscript{th}~WE-Heraeus-Seminar `Experimental Tests and Signatures of Modified and Quantum Gravity' for the opportunity to present and discuss the topics which laid the foundation for this work.
\end{acknowledgement}

\bibliographystyle{spmpsci}
\bibliography{refs}
\end{document}